\documentclass[12pt]{iopart}
\pdfoutput=1
\usepackage[pdftex]{hyperref,color,graphicx}
\usepackage[utf8]{inputenc}
\usepackage[titletoc]{appendix}

\hypersetup{
	pdfpagelabels=true,
	colorlinks=true,
	linkcolor=blue,
	anchorcolor=blue,
	menucolor=blue,
	urlcolor=blue,
	citecolor=blue,
	filecolor=blue,
	urlcolor=blue,
	pagecolor=blue,
	pdfkeywords={Quantum walks, Decoherence, Optical lattices, Floquet theory},
	pdfauthor={Alberti et al.},
	pdftitle={Decoherence Models for Discrete-Time Quantum Walks and their Application to Neutral Atom Experiments}}

\usepackage{iopams}   
\usepackage[english]{babel}
\usepackage{cite} 

\newcommand*{\ket}[1]{|{#1}\rangle}
\newcommand*{\bra}[1]{\langle{#1}|}
\newcommand{\figref}[2]{\hyperref[#1]{\ref{#1}(#2)}}
\newcommand{\figre}[1]{\hyperref[#1]{\ref{#1}}}
\renewcommand\textemdash{\leavevmode\unskip\kern0.8pt\rule[0.215\baselineskip]{8pt}{0.22pt}\kern1pt\ignorespaces}

\usepackage[separate-uncertainty=true]{siunitx}
\usepackage{hypernat}

\newif\ifusebibfile
\usebibfilefalse
\usebibfiletrue

\DeclareSIUnit\gauss{G}
\DeclareSIUnit\sqrthertz{$\sqrt{\mathrm{Hz}}$}

\makeatletter
\renewcommand\l@section[2]{%
  \ifnum \c@tocdepth >\z@
    \addpenalty{\@secpenalty}%
    \addvspace{0.2em \@plus\p@}
    \setlength\@tempdima{1.5em}%
    \begingroup
      \parindent \z@ \rightskip \@pnumwidth
      \parfillskip -\@pnumwidth
      \leavevmode \bfseries
      \advance\leftskip\@tempdima
      \hskip -\leftskip
      #1\nobreak\hfil \nobreak\hbox to\@pnumwidth{\hss #2}\par
    \endgroup
  \fi}
\makeatother

\newcommand\eprint[2][]{\href{#1}{\tt{}#2}}

\begin{document}
\selectlanguage{english}

\title{Decoherence Models for Discrete-Time Quantum Walks and their Application to Neutral Atom Experiments}
\author{Andrea Alberti$^1$, Wolfgang Alt$^1$, Reinhard Werner$^2$, Dieter Meschede$^1$}

\address{$^1$ Institut f\"ur Angewandte Physik, Universit\"at Bonn, Wegelerstra\ss{}e 8, 53115 Bonn, Germany}
\address{$^2$ Institut f\"ur Theoretische Physik, Leibniz Universit\"at Hannover, Appelstra\ss{}e 2, 30167 Hannover, Germany}
\ead{alberti@iap.uni-bonn.de}

\date{\today}


\begin{abstract}
We discuss decoherence in discrete-time quantum walks in terms of a phenomenological model that distinguishes spin and spatial decoherence.
We identify the dominating mechanisms that affect quantum walk experiments realized with neutral atoms walking in an optical lattice.

From the measured spatial distributions, we determine with good precision the amount of decoherence per step, which provides a quantitative indication of the quality of our quantum walks. 
In particular, we find that spin decoherence is the main mechanism responsible for the loss of coherence in our experiment.
We also find that the sole observation of ballistic instead of diffusive expansion in position space is not a good indicator for the range of coherent delocalization.

We provide further physical insight by distinguishing the effects of short and long time spin dephasing mechanisms.
We introduce the concept of coherence length in the discrete-time quantum walk, which quantifies the range of spatial coherences.
Unexpectedly, we find that quasi-stationary dephasing does not modify the local properties of the quantum walk, but instead affects spatial coherences.

For a visual representation of decoherence phenomena in phase space, we have developed a formalism based on a discrete analogue of the Wigner function.
We show that the effects of spin and spatial decoherence differ dramatically in momentum space.

\vspace{2mm}\noindent{\it Keywords\/}: Quantum walks, Decoherence, Optical lattices, Floquet theory
\end{abstract}

\maketitle

\tableofcontents

\section{Introduction}

Coherent superposition of quantum states constitutes the key element of every application of quantum technologies like quantum metrology, quantum communication and quantum simulation.
In real-world applications, quantum superposition states are highly fragile because they are always subject to decoherence and dephasing mechanisms.
Preventing these states from undergoing rapid loss of coherence represents today the main experimental challenge for advancing quantum technology from elementary demonstrations to more complex devices.
A detailed and quantitative understanding of decoherence is thus essential.

Over the last few years, experiments have been extending their control to increasingly larger Hilbert spaces.
In practice, such a development requires experimental access to a larger and larger number of degrees of freedom, which can be coherently manipulated and yet sufficiently decoupled from any external noise source.
Systems like trapped ions \cite{Blatt:2012} and superconducting circuits \cite{Devoret:2013} \textemdash to mention only a few of them \textemdash have been among the main driving forces of quantum technological evolution.
In the same vein, the realization of discrete-time quantum walks provides another example of coherent manipulation of a large quantum system \cite{Karski:2009}: implementing a $n$-step quantum walk indeed requires engineering and controlling a $2n$-dimensional Hilbert space.

Theoretical methods to treat decoherence in discrete-time quantum walks have been the subject of several theoretical publications \cite{Kendon:2007}.
Decoherence mechanisms like coin decoherence and spatial decoherence have been suggested, and their effect on the spatial probability distribution has been numerically calculated \cite{Dur:2002}.
Analytic solutions for the special case of spin decoherence have also been derived to give an account of the quantum-to-classical transition when decoherence sets in \cite{Brun:2003a,Annabestani:2010}.
More recently, the asymptotic spatial distributions in the presence of translationally invariant decoherence mechanisms have been analytically derived by means of a generalized group velocity operator~\cite{Ahlbrecht:2011}.
With a similar formalism based on perturbation theory, the spatial distribution in the presence of both decoherence and spatial disorder has been analytically studied~\cite{Ahlbrecht:2012}.
Decohered quantum walks have also been considered for algorithmic applications, namely for searching unstructured databases \cite{Kendon:2003}.

In this work, we use a phenomenological decoherence model to analyze the quantum walk of a single cesium atom moving along a one-dimensional optical lattice. By comparing this model with our experimental data, we retrieve information about the main physical mechanisms that are responsible for the loss of coherence (see section \ref{sec:phenomomdel}).
The agreement between the model and the experimental data allows us to infer with good precision the amount of decoherence per step \textemdash the primary figure of merit that determines the quality of a quantum walk.
As a complement to this phenomenological analysis, we provide a thorough discussion of physical decoherence mechanisms occurring in quantum walk experiments that are based on neutral atoms in an optical lattice (see section \ref{sec:decoh_mech} with related appendix \ref{sec:appendix}).
It should be noted that understanding decoherence processes in optical lattices is also relevant to a much larger class of optical lattice experiments, which hold the promise for future quantum technologies.

By analyzing the spatial distribution of quantum walks up to $40$ steps, we identify spin decoherence as the dominant mechanism in our system.
Moreover, we provide new physical insight into decoherence by analyzing the motion of the walker in Fourier space, that is, analyzing the evolution of the probability distribution of momentum in addition to the spatial distribution (see section \ref{sec:decohmomspace}). The momentum distribution exhibits striking differences between coin and spatial decoherence, in stark contrast to the position distribution, which displays only small differences between the two decoherence models.

Broome \emph{et al.}~\cite{Broome:2010} have studied decoherence in quantum walks with single
polarized photons propagating through a small series of 6 birefringent
displacers.
Misalignment of the optical components reduced the overlap between the different paths. The gradual suppression of interference resulted in a transition to the classical random walk.
In a different approach, Schreiber~\emph{et al.}~\cite{Schreiber:2011} measured the influence of decoherence on the variance of the spatial distribution by using a wave-like analogue of quantum walks \textemdash a fiber loop network hosting circulating coherent laser pulses.
An electro-optic modulator imprinted controllable phase fluctuations between vertical and horizontal polarizations, which are randomized at each position and step causing a transition to diffusive motion.
These experiments did not distinguish, though, decoherence mechanisms related to the coin and shift degrees of freedom in quantum walks.

\section{Quantum Walks in Position and Momentum Space}

We call ``quantum walker'' a quantum particle that moves in discrete steps with the step size and the direction being conditioned on the particle's internal state.
In this work, we specifically consider the quantum walk of a pseudo spin-1/2 particle \textemdash that is, of a quantum particle possessing two controllable internal states \textemdash on a periodic, one-dimensional lattice.

In such a quantum walk on the line, the walker performs repeated positional shifts to the right or to the left by one lattice site conditional on its spinor state. Before each shift, a coin operation rotates the internal spinor state on the Bloch sphere.
The state of a quantum walker is therefore represented in the product Hilbert space that combines a spin-1/2 system with a lattice of evenly spaced positions,
\begin{equation}\label{eq:states}
    \ket{s,x} = \ket{s}\otimes \ket{x}
    \quad \mbox{with} \quad s = \{\uparrow,\downarrow\}
    \quad \mathrm{and} \quad  x = 0, \pm 1, \pm 2, \dots\,.
\end{equation}
While in our laboratory quantum walks are implemented by indeed shifting the position of single atoms in an optical lattice \cite{Karski:2009,Genske:2013}, other realizations have instead employed abstract position spaces such as the time domain with circulating light pulses \cite{Schreiber:2011} or the phase space with single trapped ions \cite{Schmitz:2009,Zahringer:2010}.

\subsection{Quantum Walks in Position Space}
The state of a quantum walk after $n$ steps is obtained by repeatedly applying $n$ times the walk operator $\hat{W}$ onto the state of the quantum walker. The walk operator $\hat{W}=\hat{S}\hat{C}$ comprises two discrete operations: the conditional shift $\hat{S}$ and the coin $\hat{C}$.

The conditional shift operator
\begin{equation}\label{eq:shift}
    \hat{S} =
       \ket{\uparrow}\bra{\uparrow}\otimes\sum_x\ket{x+1}\bra{x}
       +
       \ket{\downarrow}\bra{\downarrow}\otimes\sum_x\ket{x-1}\bra{x}
\end{equation}
translates the walker to the left or to the right neighboring site
conditioned on the spin state. The coin operation $\hat{C}$ affects only the spin state rotating it by an angle $\theta$ with respect to the $y$-axis
\begin{equation}\label{eq:coin}
\hat{C} = \exp(-i \hat\sigma_y\theta/2)
\end{equation}
where $\hat\sigma_y$ is the $y$-component of the Pauli matrices $\hat{\sigma}_i$ (there is no loss of generality in assuming the rotation along a given axis on the equatorial plane of the Bloch sphere).

The coin angle $\theta$ essentially determines the unbalance of the walk. The most suited angle to study the influence of decoherence in quantum walks is  $\theta=\pi/2$, which is the corresponding analogue of the classical 50-50 fair coin. In fact, the fair-coin walk \textemdash best known as Hadamard walk \textemdash entails the largest amount of quantum superposition during its evolution \cite{robens:2014}: each step of the walk maximally mixes the two spin components in a coherent superposition. In contrast, coin angles in the proximity of $\theta=0$ or $\theta=\pi$ correspond to the opposite situation, in which a walker prepared originally in a given site and spin state follows a purely classical trajectory \textemdash thus with no superposition state involved. These quasi classical quantum walks are not affected by decoherence.

\subsection{Quantum Walks in Momentum Space}\label{sec:qwalksmomspace}
Quantum walks on a periodic lattice are invariant under discrete translations of the lattice itself.
Not only has translational invariance an important effect on the eigenstates of the walk operator $\hat{W}$ and its quasi energy spectrum, but it also has a significant bearing on how decoherence affects the walker's motion.

\begin{figure}
	\centering
	\includegraphics[width=0.9\textwidth]{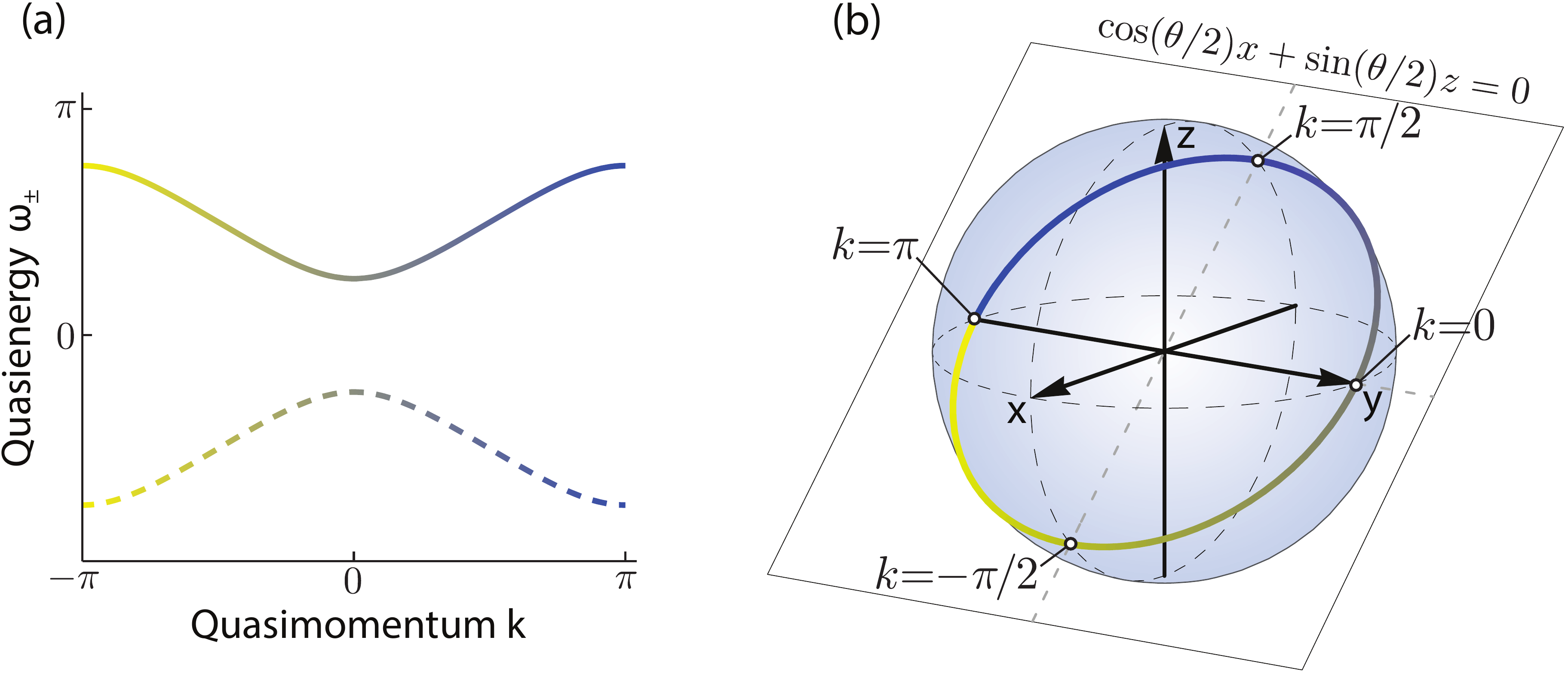}
	\caption{\label{fig:Figure1}Quasi energy spectrum and eigenspinors of a Hadamard quantum walk. (a) Solid and dashed curves represent the two energy bands according to (\ref{eq:bands}). (b)
The great circle shown on the Bloch sphere represents the ensemble of eigenspinors that are associated to the upper band in (a). The plane in which the eigenspinor rotates while sweeping across the Brillouin zone is also shown. The color gradient (from yellow to blue) matches the quasi momenta in (a) to the corresponding eigenspinors in (b).}
\end{figure}

By introducing the translation operators $\hat{T}_{\pm 1} = \exp(\mp i \hat{k})$ that translate the walker by one site either to the right or to the left, the shift operator $\hat{S}$ can simply be rewritten as
\begin{equation}\label{eq:shiftop}
	\hat{S} =
	\ket{\uparrow}\bra{\uparrow}\otimes \hat{T}_{+1}
	+
	\ket{\downarrow}\bra{\downarrow}\otimes  \hat{T}_{-1} = \exp(-i\hat{\sigma}_z \hat{k})\,,
\end{equation}
where $\hat{k}$ is the quasi momentum operator and the length unit is chosen equal to the lattice constant.
Hence, the walk operator $\hat{W}$ is diagonal in the $k$-basis and can be reduced to $\hat{W}(k)=\bra{k}\hat{W}\ket{k}$ in a subspace with given quasi momentum $\ket{k}=(2\pi)^{-1/2}\sum_x \exp(ikx)\ket{x}$. The operator $\hat{W}(k)$ acts on the spin state and its off-diagonal elements measure the spin-orbit coupling strength that is proportional to $\sin(\theta/2)$.
The diagonalization of $\hat{W}(k)$ yields the quasi energy spectrum composed of two symmetric energy bands
\begin{equation}
	\label{eq:bands}
	\omega_{\pm}(k) = \pm \arccos \left(\cos(\theta/2) \cos(k)\right)\,,
\end{equation}
which are displayed in Figure~\figref{fig:Figure1}{a}.
For each quasi momentum $k$ in the Brillouin zone, there exist two orthogonal eigenspinors, $\ket{s_{+}(k)}$ and $\ket{s_{-}(k)}$, one for each band, which are represented on the Bloch sphere by polar and azimuthal angles, respectively,
\begin{equation}\label{eq:eigenspinors}
\hspace{-20pt}	\vartheta_+=2\arctan\hspace{-2pt}\left(\frac{\tan(\theta/2)}{\sqrt{\tan^2(\theta/2)+\sin^2(k)}+\sin(k)}\right)\quad\mathrm{and}\quad\varphi_+ = \pi/2+k
\end{equation}
for the upper band, and $\vartheta_-=\pi-\vartheta_+$ and $\varphi_-=\varphi_++\pi$ for the lower band, as shown in Figure~\figref{fig:Figure1}{b}.
The spin-orbit interaction causes a gap both at $k=0$ and $k=\pi/d$, which separates the two energy bands by an amount equal to $\theta$. Note that, in the spirit of the Floquet theory, the dimensionless spectrum $\omega_\pm(k)$ can, if necessary, be recast in physical units by multiplying it by $\hbar/\tau$, where $\tau$ is the real duration of a single step of the walk.

\section{Ballistic Expansion of Long Quantum Walks of Single Atoms}\label{sec:40steps}

A quantum walker that is initially localized on a single lattice site is in an equal superposition of all quasi momentum states, although the distribution between the two bands, in general, still depends on $k$.
In the absence of decoherence, the size of the wave packet after $n$ steps is simply obtained by averaging the group velocity (expressed in units of sites per step)
\begin{equation}\label{eq:groupvelocity}
	v_{\text{g},\pm}(k)=\partial\omega_\pm(k)/\partial k
\end{equation}
over the entire Brillouin zone \cite{Ahlbrecht:2011},
\begin{equation}\label{eq:sizewavepacket}
	\langle \Delta x^2\rangle = n^2 \hspace{-2pt} \int_{-\pi}^{+\pi}\frac{\mathrm{d}k}{2\pi}\, v_{\hspace{-1pt}g\hspace{1pt}\pm}(k)^2= n^2\left(1-|\sin(\theta/2)|\right).
\end{equation}
In contrast to classic random walks exhibiting diffusive spreading
of an initially localized particle ($\langle\Delta x^2\rangle = {n}$), equation (\ref{eq:sizewavepacket}) shows that the state of the walker is delocalized over a number of sites that scales linearly with $n$. The wave packet performs a ballistic expansion, which is analogous to that observed in continuous-time system \cite{Ivanov:2008}.

\begin{figure}
	\centering
	\includegraphics[width=0.95\textwidth]{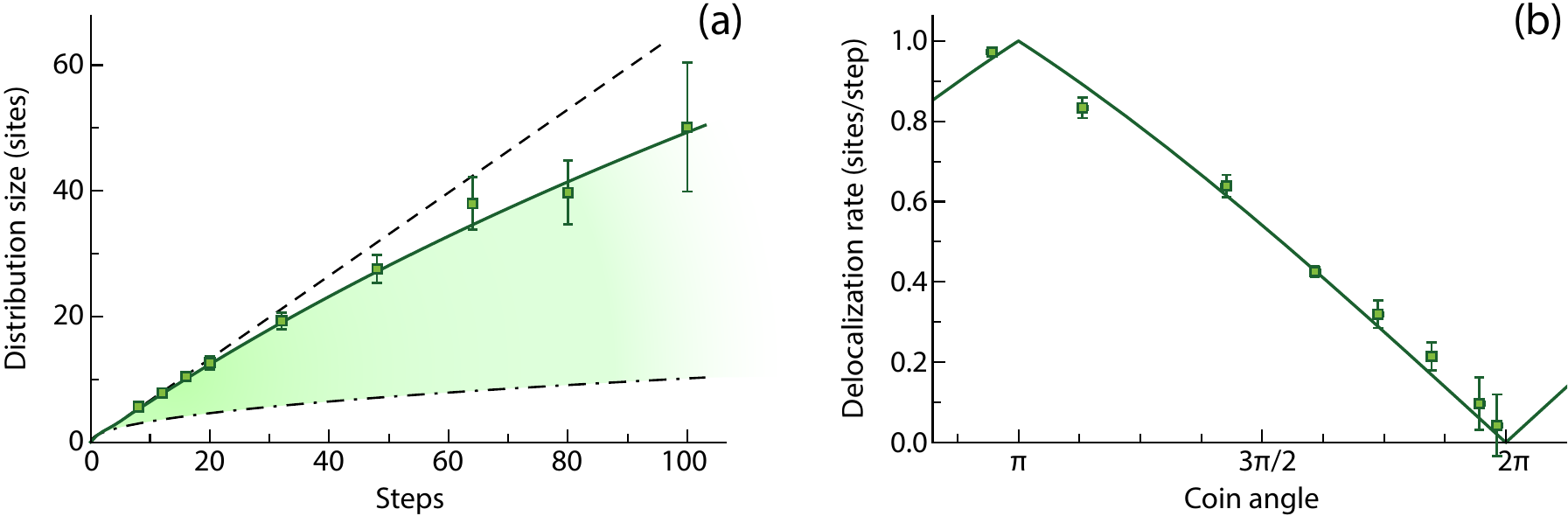}
	\caption{\label{fig:Figure2}
Coherent delocalization of single atoms during a one-dimensional discrete-time quantum walk. (a) Measured root-mean-square (RMS) width $\langle \Delta x^2\rangle$ of the spatial distribution of quantum walks for an increasing number of steps (square points with $1\,\sigma$ error bars). The coin angle is set close to $3\pi/2$, precisely $\theta\approx 1.38\,\pi$.
The curves represent the theoretical predictions for an ideal quantum walk (dashed), for a quantum walk with spin dephasing of $\SI{5}{\percent}$ per step (solid), and for a classical random walk (the dash-dotted).
(b) Measured delocalization rate as a function of the coin angle $\theta$  (data points with error bars). The points are obtained from the RMS width of a 12-step quantum walk normalized to the total number of steps. The solid curve displays the theoretical prediction of equation (\ref{eq:sizewavepacket}) after mapping $\theta\rightarrow\theta+\pi$ to account for the different definition of the coin angle used in the experiment.
}
\end{figure}

In the experiments, we interpret the transition from ballistic to diffusive behavior occurring at a large number of steps as the onset of decoherence effects.
Figure~\figref{fig:Figure2}{a} shows that this transition occurs at the present in our experiment after about 40 steps, while initial experiments deviated from ballistic spreading already after about 10 steps \cite{Karski:2009}.
The experimental data have been acquired with the same experimental apparatus employed in \cite{Genske:2013}:
Individual cesium atoms are initialized in a single site ($x{=}0$ position) of an optical lattice (lattice constant $\SI{433}{\nano\meter}$).
To suppress inhomogeneous dephasing due to thermal motion, we cool the atoms along the lattice direction to the lowest vibrational state~\cite{Belmechri:2013}.
Microwave pulses are used both to prepare the initial spin state and to perform the coin operation in Eq.~(\ref{eq:coin}) coupling the two hyperfine states $\ket{\downarrow}=\ket{F=3,m_F=3}$ and $\ket{\uparrow}=\ket{F=4,m_F=4}$ of the electronic ground state.
According to the original idea proposed by D\"ur \emph{et al.}~\cite{Dur:2002}, the state-dependent shift operation in Eq.~(\ref{eq:shiftop}) is performed by controlling the polarization of the laser beams forming the optical lattice by means of an electro-optic modulator \cite{Belmechri:2013}.
However, instead of applying $\hat{S}$ on each step, we rather alternate  $\hat{S}$ and $\hat{S}^\dagger$ by a applying a sawtooth-like signal to the electro-optic modulator.
Simple algebra using Pauli matrices reveals that the experiment implements (up to a global phase change and a spin flip of the final state) a Hadamard walk with a redefined coin angle $\theta \rightarrow \theta +\pi$ and a spin flip of the initial state.

Hence, the spreading behavior of the walker provides us with a first quantitative estimate of the degree of quantumness of the walk.
However, such an analysis does not allow us to discriminate between different physical decoherence mechanisms.
In addition, one should also be aware that the spreading velocity strongly depends on the coin angle both in quantum walks and in classical random walks.
Figure~\figref{fig:Figure2}{b} shows the measured delocalization rates of a quantum walk for different coin angles.
In general, the observation of a large spreading speed does not strictly imply good coherence properties since, for instance, a walk with no coin operation ($\theta=\pi$ in the figure) always spreads out ballistically regardless of the amount of decoherence.
Related findings have experimentally been obtained by analyzing slow fluctuations in the wave-mechanics-analogue implementation of quantum walks~\cite{Schreiber:2011}.

Information about which type of coherence is preserved or lost \textemdash a relevant question for any quantum technology experiment \textemdash can only be obtained through a detailed analysis of the probability distribution of the walk.
In our experiment, the probability distribution in position space can directly be reconstructed by fluorescence imaging of atoms' positions, whereas the distribution in momentum space is not easily accessible\cite{note:TOF}.
The question addressed in this work is how physical information about decoherence can be retrieved from small, though distinctive features, which we have been systematically observing in the spatial probability distribution.
For instance, we always record a cusp-like maximum of the spatial distribution at the $x=0$ position, which is not predicted by the decoherence-free theory~\cite{Karski:2009}.
We tackle this question by employing a phenomenological model that assigns decoherence rates separately to the two degrees of freedom: spin and position.

\section{Phenomenological Model of Memoryless Dephasing Processes}\label{sec:phenomomdel}

\subsection{Spin and Spatial Decoherence Model}\label{sec:spinspatdecohmodel}
We begin with a conventional approach based on the density operator formalism to describe the non-unitary evolution of the quantum walk.
The density operator is given by
\begin{equation}\label{eq:density}
  \hat{\rho} =\hspace{-6pt} \sum_{x,x';s,s'}\,
	 \hspace{-4pt}\rho_{x,x';s,s'}\hspace{2pt}\ket{x,s}\bra{x',s'}
\end{equation}
where the matrix $\rho_{x,x';s,s'}$ contains the probabilities of occupying some given site and spin state (diagonal terms) and the amount of coherence between quantum states $|x,s\rangle$ and $|x',s'\rangle$ (off-diagonal terms).
With no decoherence, the evolution of the density operator from step $n$ to $n+1$ is simply represented by the unitary process
\begin{equation}\label{eq:unievol}
  \hat{\rho}(n+1) = \hat{W}\hat{\rho}(n)\hat{W}^\dagger,
\end{equation}
where $\hat\rho(n)$ describes the states after $n$ steps.
By including decoherence, instead, the evolution of the walk becomes non-unitary.
We assume a simple model in which decoherence takes place at the end of each step in a discrete-time manner.
This assumption is appropriate in experiments \textemdash in particular ours \textemdash in which the walker's evolution over a single step is predominantly coherent.
With this model, we replace the continuous decoherence and dephasing processes, which occur during the entire duration of each step, with a single discrete operation.

In particular, we consider two classes of decoherence that cause the relaxation of spin coherences and spatial coherences, respectively.
We do not consider, instead, relaxation of populations, which can be caused, for instance, by spin flips or tunneling into adjacent lattice sites, because these effects take place on a much longer timescale under typical experimental conditions.
On each step, spin decoherence leaves the spin-diagonal elements $\rho_{x,x';s,s}$ unchanged, while it reduces the spin-off-diagonal elements $\rho_{x,x';s\neq s'}$ by a factor $1{-}p_\text{C}$. Similarly, spatial decoherence preserves the position-diagonal elements $\rho_{x,x;s,s'}$ while it suppresses on each step the position-off-diagonal elements $\rho_{x\neq x';s,s'}$ by a factor $1{-}p_\text{S}$.
This assumption is well suited to describe memoryless decoherence mechanisms such as, for example, dephasing produced by an external field with white spectral noise.
Although in principle all physical decoherence mechanisms possess a finite correlation time (memory), in practice the ``coarse-grained'' discretization in time of the quantum walk's evolution allows us to model the loss of coherence with some fixed decay rates $1-p_\text{C}$ and $1-p_\text{S}$.
A model of decoherence with long-time memory is provided in Section~\ref{sec:long-time-memory}.

Hence, the quantum operation that represents the single step of the walk can be decomposed into a unitary, decoherence-free contribution (with probability ${1{-}p_\text{C}{-}p_\text{S}}$) and an ensemble of non-unitary contributions, which project the walker (namely, the two-level atom) either into a definite spin state $s=\{\uparrow,\downarrow\}$ or into some given lattice site $x$ with probabilities $p_\text{C}$ and $p_\text{S}$, respectively:
\begin{eqnarray}\label{eq:nonunievol}
	\hat{\rho}(n+1) =&  (1-p_\text{C}-p_\text{S})\,\hat{W}\hat{\rho}(n)\hat{W}^\dagger \\[1.5mm]
	&+	p_\text{C} \sum_s \hat{\mathbb{P}}_s\hat{W}\hat{\rho}(n)\hat{W}^\dagger\hat{\mathbb{P}}_s^\dagger
	+ p_\text{S} \sum_x \hat{\mathbb{P}}_x\hat{W}\hat{\rho}(n)\hat{W}^\dagger \hat{\mathbb{P}}_x^\dagger\,,\nonumber
\end{eqnarray}
where the operator $\hat{\mathbb{P}}_s=\sum_{x}\ket{x,s}\bra{x,s}$ projects the spin state onto $\ket{s}$ and the operator $\hat{\mathbb{P}}_x=\sum_{s}\ket{x,s}\bra{x,s}$ projects the position state onto $\ket{x}$.
The process (\ref{eq:nonunievol}) has a self-evident operator-sum representation $\hat{\rho}(n+1) = \sum_k \hat{E}_k\hspace{1pt}\hat{\rho}(n)\hspace{1pt}\hat{E}_k^\dagger$ in terms of the Kraus operators
\begin{equation}\label{eq:krausops}
	\hat{E}_0 = \sqrt{1-p_\text{C}-p_\text{S}}\hspace{2pt} \hat{W},\quad 	\hat{E}_s = \sqrt{p_\text{C}}\hspace{3pt}\hat{\mathbb{P}}_s\hat{W},\quad   	\hat{E}_x = \sqrt{p_\text{S}}\hspace{3pt}\hat{\mathbb{P}}_x\hat{W},
\end{equation}
with $\hat{E}_0$ being responsible for the coherent evolution and the ensemble of $\hat{E}_s$ and $\hat{E}_x$ operators contributing to the damping of the off-diagonal terms in $\hat{\rho}$.

Specifically for the spin decoherence model, Brun \emph{el al.}\ derived the analytic expression of the first two moments of the quantum walk's spatial distribution \cite{Brun:2003a},
finding that the quantum walk's behavior turns asymptotically into diffusive spreading with the variance equal to $\langle \Delta x^2\rangle = D\hspace{1pt}n$ (up to an additive constant).
The diffusion constant $D={[{1+(1-p_\text{C})^2}]/[{1-(1-p_\text{C})^2}]}$ diverges for the coherent quantum walk ($p_\text{C}=0$), while it becomes exactly $1$ for the fully incoherent walk ($p_\text{C}=1$) \textemdash namely, the same diffusion constant of the classical random walk.
The authors suggest that the match of the diffusion constant with that of a classical random walk provides evidence of the full quantum-to-classical transition.
On the contrary, the analysis of spin decoherence in Section~\ref{sec:decohmomspace} reveals that states with defined quasi momentum preserve spatial coherence even for $p_\text{C}=1$.

\subsection{Density Matrix and Spatial Coherences}

\begin{figure*}[b]
	\centering
		\includegraphics[width=.9\textwidth]{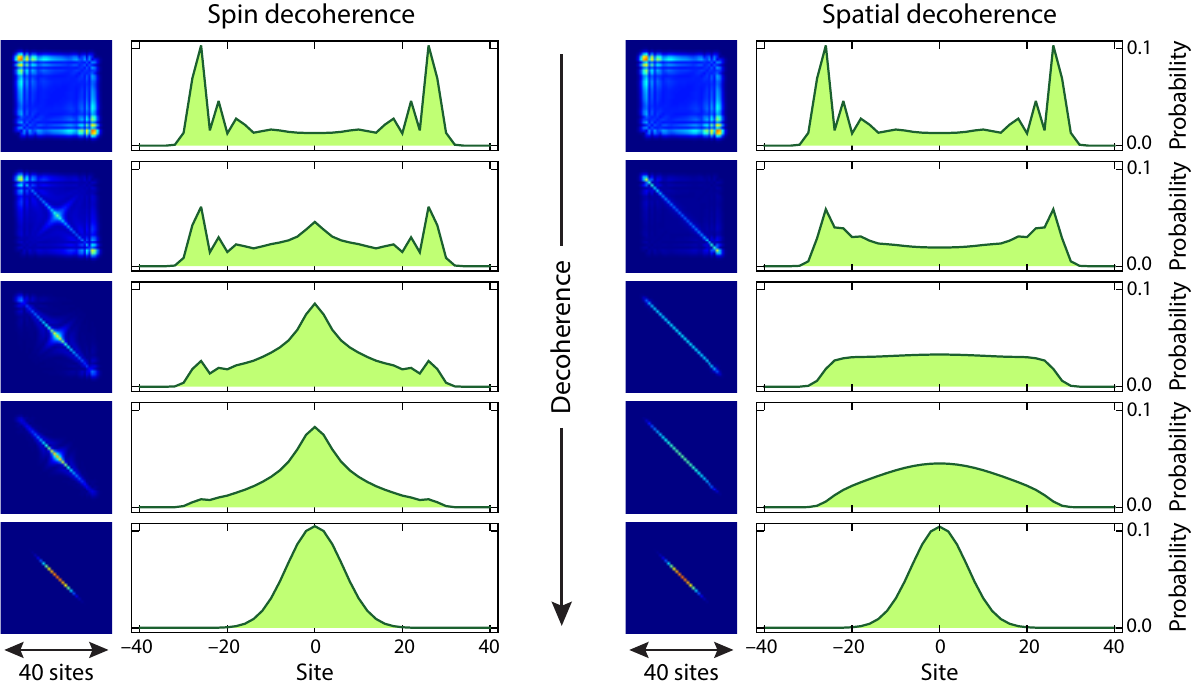}
	\caption{Theoretical spatial density matrix and probability distributions of a 40 step Hadamard quantum walk in the presence of decoherence. The walker starts in the localized, spin-symmetric state $(\ket{\uparrow,0}+i\ket{\downarrow,0})/\sqrt{2}$. Only even sites are displayed because the probability to occupy odd sites is equal zero. Left and right columns describe pure spin and spatial decoherence, respectively. The amount of decoherence increases from top to bottom taking the values: $p_\text{C,S}=\num{0.00}$, $\num{0.05}$, $\num{0.15}$, $\num{0.25}$ and $\num{1.00}$. The square panels display in false color scale the absolute value of the density matrix traced over the spin degree of freedom (values increase from dark blue to red).
}
	\label{fig:Figure3}
\end{figure*}

The spatial probability distribution is admittedly the measurable physical quantity that is most often reported when dealing with quantum walks: it can be easily visualized and directly accessed in experiments, it has a direct correspondence with the probability distribution of a classical random walk, and its expansion properties lie at the heart of quantum search algorithms.
However, the spatial probability distribution provides only limited information. For a deeper understanding of the evolution of a (decohered) quantum walk, we should rather examine the entire physical information stored into the density matrix.

We illustrate the effect of decoherence on the spatial coherences of a quantum walk by letting only one degree of freedom (either spin or position) relax.
In Figure~\ref{fig:Figure3}, we show the result of a numerical evaluation of the density matrix elements.
From top to bottom, we have increased the decoherence rates $p_\text{C,S}$ with selected values ranging from 0 (no decoherence) to 1 (fully decohered).
In displaying the density matrix, we trace out the spin components, $\mathrm{Tr}_\textrm{S}(\hat{\rho}) {=} \sum_{s}\rho_{x,x';s,s}$, to avoid treating each spin-spin combination separately.
The top-left to bottom-right diagonal terms $\bra{x}\mathrm{Tr}_\textrm{S}(\hat{\rho})\ket{x}$ of the density matrix thus give the site occupation probabilities shown in the figure.

The comparison between spatial and spin decoherence reveals that both fully decohered walks, $p_\text{S}{=}1$ and $p_\text{C}{=}1$, yield the same density matrix with nonzero values only on the main diagonal.
The Gaussian probability distribution of a classical, diffusive random walk is thus recovered since quantum paths are not allowed to interfere in either case.
However, the transformation of a quantum walk into a classical random walk follows two qualitatively different routes for spin and spatial decoherence:
Spin decoherence causes a cusp-like maximum at the ${x{=}0}$ position in the probability distribution (see second and third raws in the figure), which is in stark contrast to the flat profile caused by spatial decoherence.
The two different behaviors have already been recognized in previous theoretical work~\cite{Kendon:2003}. New physical insight about their origin is provided in Section~\ref{sec:decohmomspace}.

\begin{figure}
	\centering
	\includegraphics[width=0.75\textwidth]{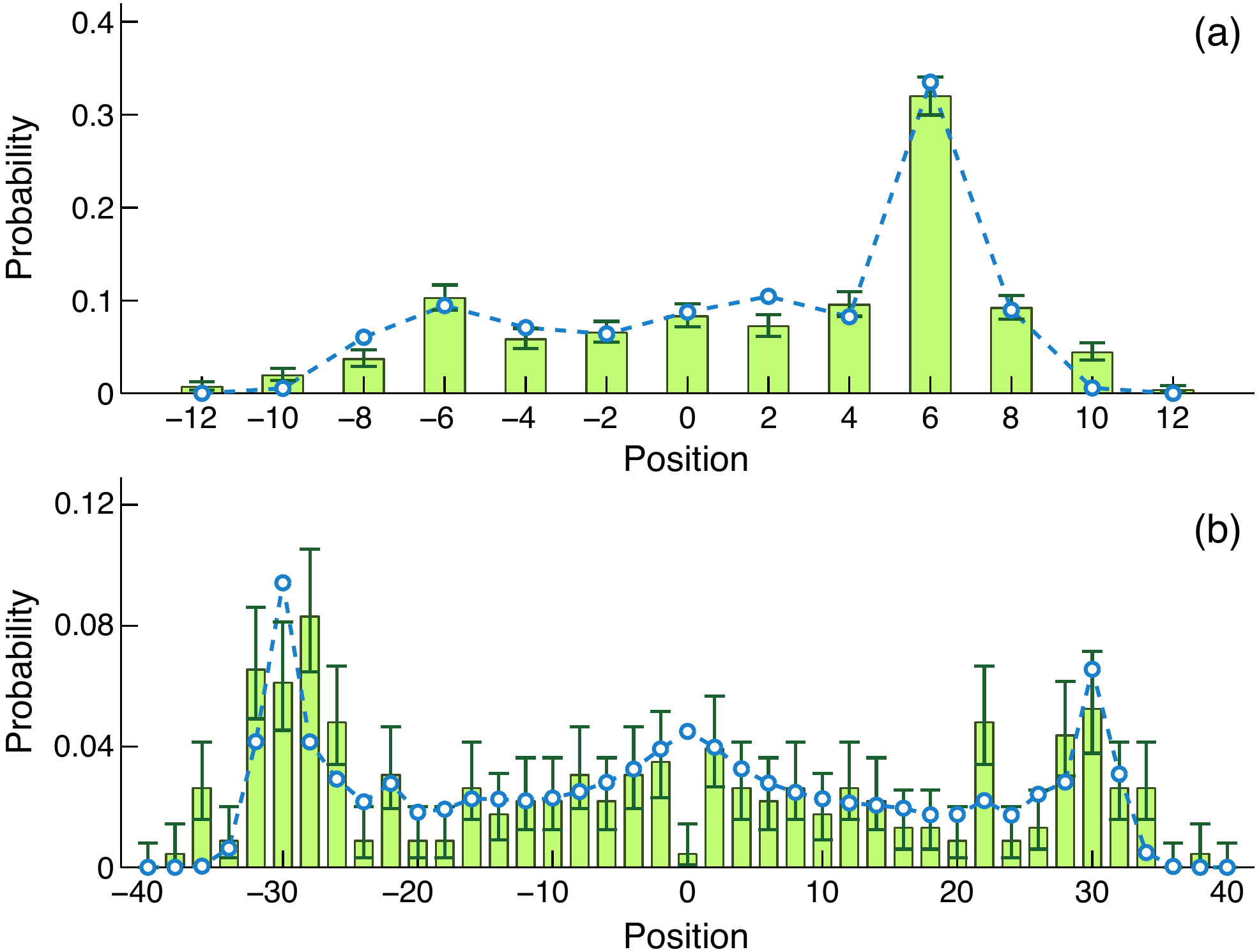}
	\caption{\label{fig:Figure4}
Measured spatial probability distribution of a
Hadamard quantum walk for (a) 12 steps and initial
$\ket{\uparrow,0}$ state, and (b) 40 steps and initial state
$(\ket{\uparrow,0}+i\ket{\downarrow,0})/\sqrt{2}$.
A pure spin decoherence model (circles) is fitted to the measured data (histogram with $\SI{68}{\percent}$ confidence Clopper-Pearson error bars).
The fit also accounts for the finite $\SI{90}{\percent}$ efficiency of our apparatus to correctly detect the discrete displacement of the atom along the lattice.
The fits yield (a) $p_\text{C}{=}\SI{1+-1}{\percent}$ under optimal experimental conditions and (b) $p_\text{C}{=}\SI{6 +- 1}{\percent}$ under slightly worse experimental conditions.
We interpret the central point in (b), which visibly deviates from the expected value, as an outlying event (occurrence probability around $\SI{0.1}{\percent}$).
}
\end{figure}

Figure~\ref{fig:Figure4} shows an application of the phenomenological decoherence model to two measured quantum walk distributions of 12 and 40 steps, respectively.
Especially for the longer walk, the distribution clearly exhibits in the proximity of the center a pronounced maximum, which we thus interpret as evidence of decoherence mechanisms affecting predominantly the spin degree of freedom.
As shown in the figure, we find a very good agreement between the experimental data and theoretical model, which has been fitted using the maximum-likelihood estimator with the coin angle $\theta$ and spin decoherence rate $p_\text{C}$ as free parameters. The spatial decoherence rate is simply fixed equal to zero, $p_\text{S}{=}0$.
Typical fitted values of $p_\text{C}$ lie around $\SI{5}{\percent}$, which matches the estimate obtained with the analysis of the RMS size of the quantum walk distributions in Figure~\ref{fig:Figure2}.
Our present finding that single atoms in state-dependent optical lattices are mainly susceptible to spin decoherence is also supported by earlier investigations, which we performed using atom interferometry techniques~\cite{Steffen:2012}.
Section~\ref{sec:decoh_mech} discusses the most relevant physical decoherence mechanisms in our apparatus.
Their detailed experimental identification and elimination requires further work.

\begin{figure*}[t]
	\centering
		\includegraphics[width=1\textwidth]{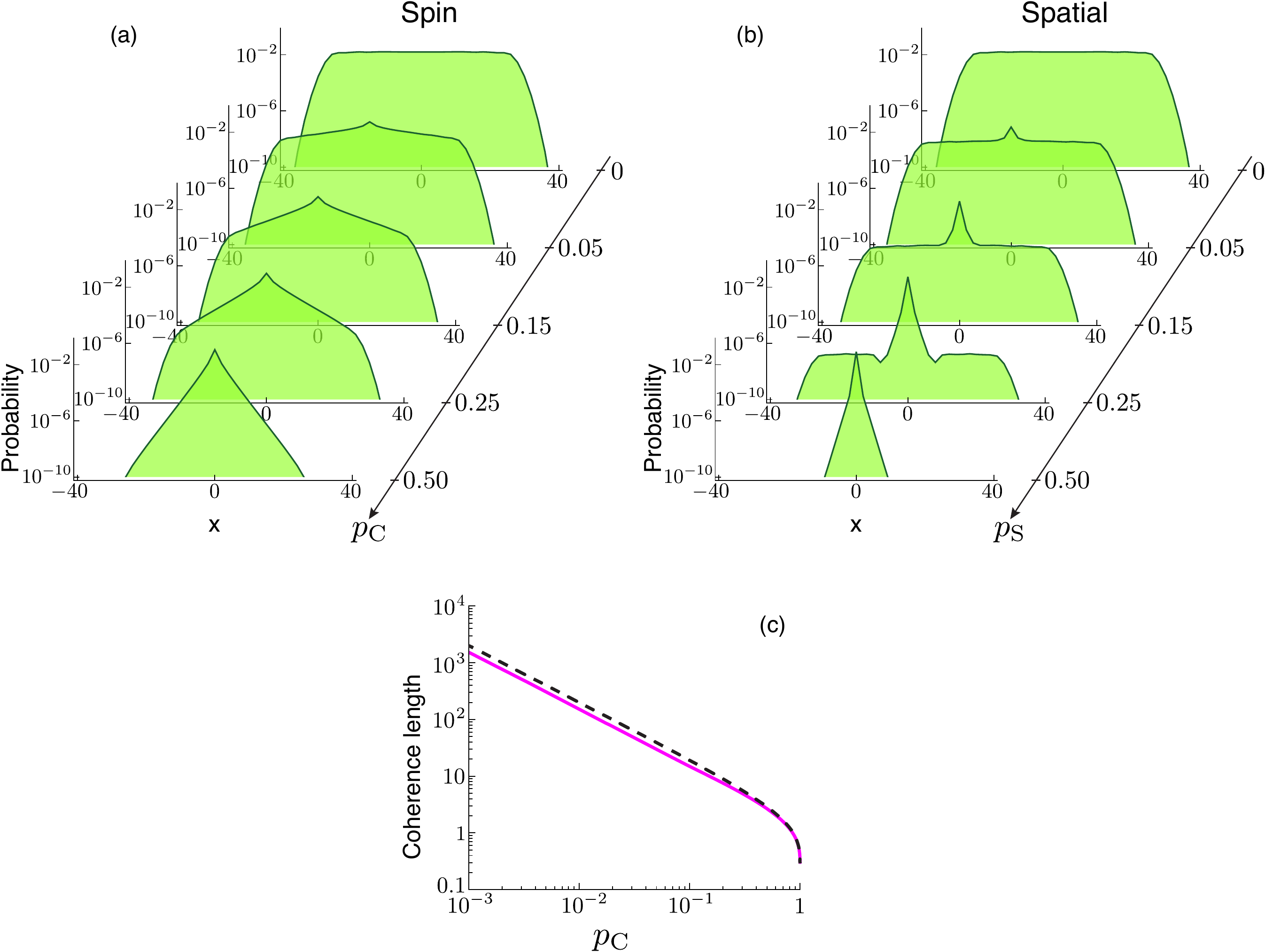}
	\caption{Single particle correlation function $|G(x,-x)|$ computed for a 40-step Hadamard quantum walk under identical conditions as those in Figure~\ref{fig:Figure3} for (a) pure spin decoherence and (b) pure spatial decoherence. (c) For pure spin decoherence, the coherence length is shown as a function of the spin decoherence rate $p_\text{C}$. The solid line is the result of a numerical analysis, the dashed line represents the simple analytical model described in the text.
}
	\label{fig:Figure5}
\end{figure*}

While the diagonal terms of $\hat{\rho}$ yield the probability distribution, the off-diagonal terms reveal the coherences of the quantum walk.
Among the off-diagonal terms, we specifically consider a particular cut through the density matrix consisting of the antidiagonal terms ${G}(x,-x)=\bra{-x}\mathrm{Tr}_\textrm{S}(\hat{\rho})\ket{x}$,
where
\begin{equation}\label{eq:singlepartcorrfunc}
G(x,y)=\sum_{s}\mathrm{Tr}(a_{x,s}^\dagger a_{y,s}\hat{\rho})	
\end{equation}
corresponds to the single particle correlation function ($a^\dagger_{x,s}$, $a_{x,s}$ create and annihilate a particle in site $x$ and spin $s$.)
The result is displayed in Figure~\ref{fig:Figure5} for different amounts of decoherence.
For a quantum walk with no decoherence, $p_\text{S}{=}p_\text{C}{=}0$, the quantum walker is maximally delocalized and spatial coherences extend rather uniformly over the entire region of populated lattice sites (flat-top profile in the figure).
With the onset of decoherence, instead, spatial coherences loose the long range character.

The effect of spin decoherence on spatial coherences is mediated by the spin-orbit coupling occurring during the shift operation $\hat{S}$.
The behavior of the correlation function in Figure~\figref{fig:Figure5}{a} seems to indicate that spatial coherences are exponentially suppressed with the distance.
By assuming, therefore, an exponential suppression of the correlation function, $|{G}(x,y)|\approx |G_0(x,y)|\exp[-{(x-y)}/\ell]$, with respect to the unperturbed one, $G_0(x,y)$, we derive the coherence length $\ell$ by means of a numerical fit procedure for different values of $p_\text{C}$.
The result, which is shown in Figure~\figref{fig:Figure5}{c}, can be compared with the estimate of $\ell$ based on a simple model:
The quantum walker must be able to perform at least $x$ steps coherently in order to develop spatial coherences between the pair of sites at positions $x$ and $-x$.
The probability for $x$ coherent steps is $(1-p_\text{C})^x$, which we equate to the suppression factor $\exp(-2x/\ell)$ of the correlation function.
We therefore obtain $\ell\approx 1/\log(1/\sqrt{1-p_\text{C}})$.
This simple model is able to reproduce very accurately the numerical result for nearly incoherent quantum walks when $p_\text{C}\lesssim1$, and it provides the correct qualitative behavior for $p_\text{C}\ll 1$, in which case $\ell$ is inversely proportional to $p_\text{C}$.

Spatial decoherence, instead, suppresses directly spatial coherences, which are uniformly reduced at each step by a factor $1{-}p_\text{S}$, as shown in Figure~\figref{fig:Figure5}{b}.
A central peak becomes dominant at larger values of $p_\text{S}$, or after a sufficiently large number of steps.
The numerical analysis of the central peak hints at an exponential fall-off of the spatial coherences with a coherence length $\ell\approx1$ for mild decoherence, that is, when $p_\text{S}$ is only a small fraction of $1$. For strong decoherence, that is, when $p_\text{S}$ approaches $1$, the coherence length becomes shorter, $\ell<1$, and eventually reaches $0$ for $p_\text{S}{=}0$.
In general, this behavior of the central peak is also found in continuous-time quantum-tunneling systems in the presence of spatial decoherence and it is thus not the prerogative of the coined quantum walk system \cite{note:continuousWalks}.

By comparing the two types of decoherence, Figure~\ref{fig:Figure5} shows that spin decoherence has significantly less influence on spatial coherences for the same decoherence rate.
In special cases, spatial coherences can even survive maximal spin decoherence $p_\text{C}{=}1$ if the initial state has a definite quasi momentum $k$, as explained in greater detail in the Section~\ref{sec:decohmomspace}.

\subsection{Wigner Function and Quantum Walks}\label{sec:wignerfnc}

The Wigner function offers an alternative way of presenting the evolution of quantum coherences during the quantum walk~\cite{Haroche:2006}.
Like the density matrix, the Wigner function contains all physical information about the quantum state of the walker.
In addition to that, however, the Wigner function provides a direct connection to the phase space structure as well, where the effects of decoherence on the quantum transport properties can be best comprehended.

The Wigner function in discrete systems has hitherto found little application  compared to its broad use in continuous phase space systems.
In discrete systems, the usual definition, which is based on the Fourier transform of the off-diagonal density matrix elements, needs to be adapted in order to account for discrete positions ($x$-space) and for cyclic boundary conditions in the Brillouin zone ($k$-space).
The proper definition that is suited for a lattice system is provided by the rotational Wigner function, which has been introduced for the equivalent phase space of the conjugated variables angle (continuous, periodic) and angular momentum (quantized)~\cite{Berry:1977,Bizarro:1994}.
We therefore obtain the definition of the Wigner function
\begin{equation}
    W_{s,s'}(x,k) = \frac{1}{\pi}\int_{-\pi/2}^{+\pi/2}\hspace{-4pt}\mathrm{d}k'
		e^{-i 2 x k'}\bra{k-k',s}{\hspace{1pt}\hat{\rho}\hspace{1pt}}\ket{k+k',s'}
\end{equation}
for the pair of $\ket{s}$ and $\ket{s'}$ spin states. The argument of the discrete Wigner function $x$ takes integer values, while $k$ takes any real value within the Brillouin zone.
We recall that the normalization condition is $\langle k \ket{k'} = \delta(k-k')$ for the quasi momentum states.
The ``marginals'' of the Wigner function yield
the probability distributions in position space, $\int_{-\infty}^{\infty}\mathrm{d}k \hspace{2pt} W_{s,s}(x,k)$, and the one in momentum space, $\sum_{x}W_{s,s}(x,k)$ for a given spin state $\ket{s}$. It holds the obvious condition $\sum_{x,s}\int_{-\infty}^{\infty}\mathrm{d}k \hspace{2pt} W_{s,s}(x)=1$ on the total probability.
A different definition of the Wigner function in a lattice system for spinless and spinor particles has recently been proposed~\cite{Hinarejos:2012,Hinarejos:2013}.
However, such an alternative definition leads to an ambiguous interpretation of the displayed results, mainly because the marginals do not coincide with the probability distributions and because of the presence of ``ghost'' image artifacts \cite{Bizarro:2013}.

\begin{figure*}[b]
	\centering
		\includegraphics[width=0.8\textwidth]{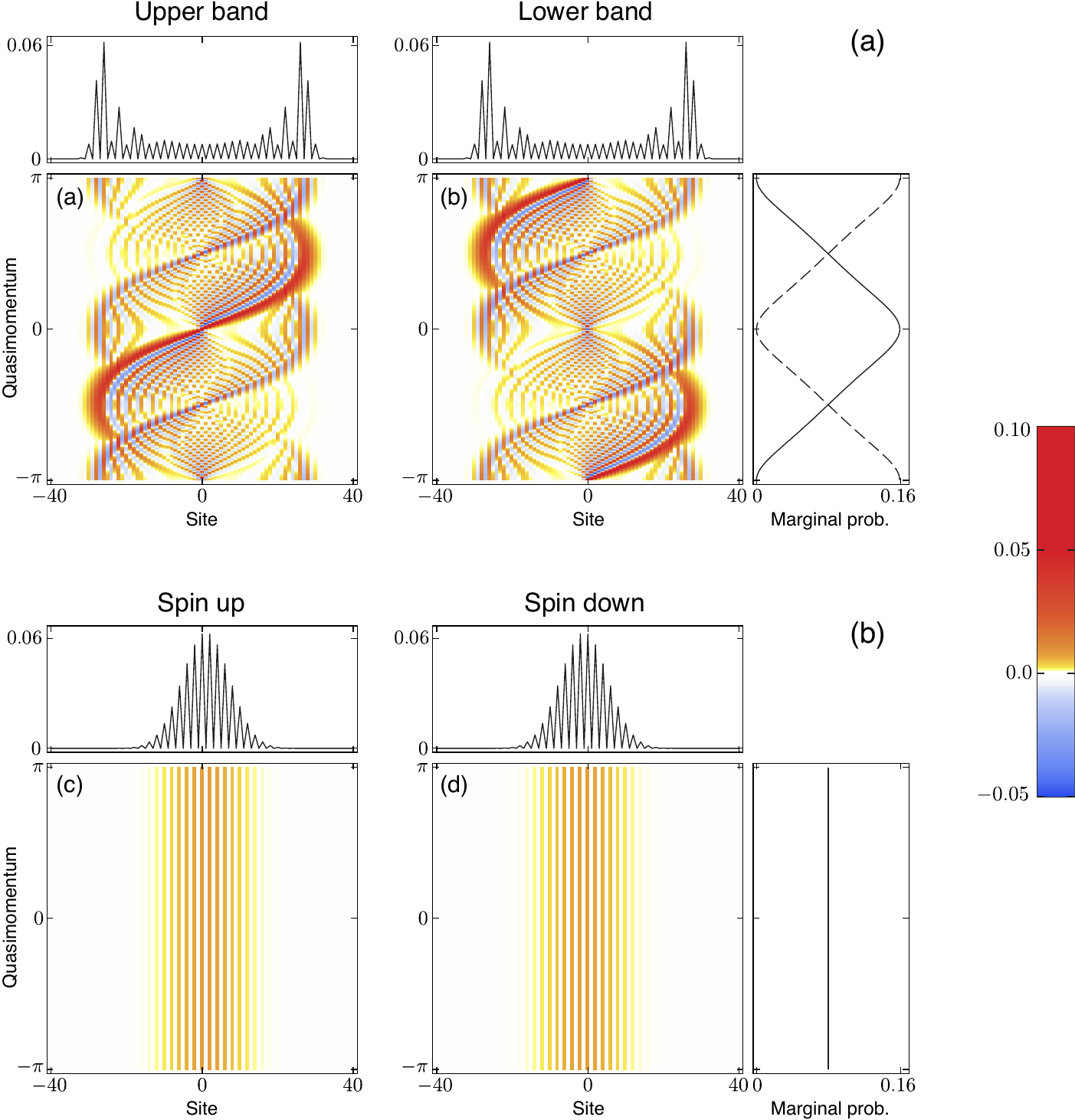}
	\caption{Phase space analysis of discrete-time quantum walks. (a) Wigner functions $W_\pm(x,k)$ computed for a 40-step Hadamard quantum walk with initial state $(\ket{\uparrow,0}+i\ket{\downarrow,0})/\sqrt{2}$ and (b) Wigner functions $W_{s,s}(x,k)$ for the corresponding classical random walk. The upper panels display the position marginal distribution. The side panels, instead, show the momentum marginal distribution, where the solid and dashed lines refer to the Wigner function on the left and right hand side, respectively (in (b) the two lines perfectly overlap).}
	\label{fig:Figure6}
\end{figure*}

Because the eigenstates have a definite spin for each quasi momentum $k$ (see Section~\ref{sec:qwalksmomspace}), it is natural to express the Wigner function in the $k$-dependent eigenspinor basis rather than in the $\{\uparrow,\downarrow\}$ spin basis.
We therefore introduce the Wigner functions
\begin{equation}
	\begin{array}{lcl}
	W_+(x,k) &=& \sum_{s,s'} \bra{s'}{s_{+}(k)}\rangle\bra{s_{+}(k)}s\rangle\,W_{s,s'}(x,k)\\[2mm]
	W_-(x,k) &=& \sum_{s,s'} \bra{s'}{s_{-}(k)}\rangle\bra{s_{-}(k)}s\rangle\,W_{s,s'}(x,k)
	\end{array}
\end{equation}
associated with the upper and lower bands, respectively.
We recall that $\ket{s_{+}(k)}$ and $\ket{s_{-}(k)}$ are the eigenspinors corresponding to the upper and lower band with quasi momentum $\ket{k}$, see the definition in Equation~\ref{eq:eigenspinors}.
The marginal distribution obtained summing over positions, $\sum_{x}W_{\pm}(x,k)$, yields the probability of occupying an eigenstate of quasi momentum $k$ in either the upper ($+$) or the lower band ($-$).
The other marginal distribution obtained integrating over quasi momenta, $\int_{-\infty}^{\infty}\mathrm{d}k \hspace{2pt}W_\pm(x,k)$, requires instead a more careful interpretation since it can take negative values. It does not represent, in fact, any probability distribution associated to a measurable observable as it is not possible (owing to the non commutativity of position and momentum) to simultaneously determine in which lattice site $x$ and in which band $\{+,-\}$ the walker resides.
Hence, only after tracing out the band index (or equivalently, the spin) we obtain the spatial probability distribution of the walker, $\sum_{\pm} \int_{-\infty}^{\infty}\mathrm{d}k \hspace{2pt} W_\pm(x,k)$.

The resulting Wigner distribution is shown in Figure~\ref{fig:Figure6} for an ideal Hadamard quantum walk after $\num{40}$ steps as well for a classical random walk ($p_
\text{C}{=}p_\text{S}{=}1$) with a fair coin toss.
The interpretation of the displayed phase space distribution merits a detailed discussion:

\begin{itemize}
	\item{}
For both walks, all quasi momenta in the Brillouin zone are uniformly populated, but with a significant difference: For the quantum walk, the quasi momentum states are in a quantum superposition, while for the random walk they instead form an incoherent statistical mixture.
This difference is evidenced by the existence of negative values of the Wigner function in the quantum walk case (blue stripes in the figure), which instead do not appear in the random walk case, where the Wigner function is always positive.
\item{}
In the absence of decoherence, the momentum probability distribution remains constant for the entire duration of the walk because the eigenstates of $\hat{W}$ possess definite quasi momentum.
If the walker is initially localized in a single lattice site, the filling of the upper and lower band depends on $k$ according to $|\bra{s_{+}(k)}\sigma\rangle|^2$ and $|\bra{s_{-}(k)}\sigma\rangle|^2$ momentum distributions, with $\ket{\sigma}$ being the initial spin state, as shown by the momentum marginal distribution in the side panel of Figure~\ref{fig:Figure6}.
\item{}
The $S$-shaped structures noticeable in the quantum walk's Wigner function reflect the behavior of the group velocity $v_{\text{g},\pm}(k)$ introduced in Equation~\ref{eq:groupvelocity}. In particular, the caustic-like distribution of the Wigner function around $k{=}\pm\pi/2$ accounts for the two outbound crests in the ballistic expansion of the quantum walk.
\end{itemize}

\subsection{Decoherence Analysis in Momentum Space}
\label{sec:decohmomspace}

We gain further insight into the origin of the cusp-like peak shown in Figure~\ref{fig:Figure4} by investigating the effect of decoherence on selected eigenstates of the walk operator $\hat{W}$.
We consider two distinct, limiting cases, in which the quantum walker is initially set either at rest ($k=0$) or in a superposition of opposite quasi momentum states $k=\pm\pi/2$ of the same band. We call the latter a $k$-cat state \cite{Monroe:1996}.

\begin{figure*}[b]
	\centering
		\includegraphics[width=.95\textwidth]{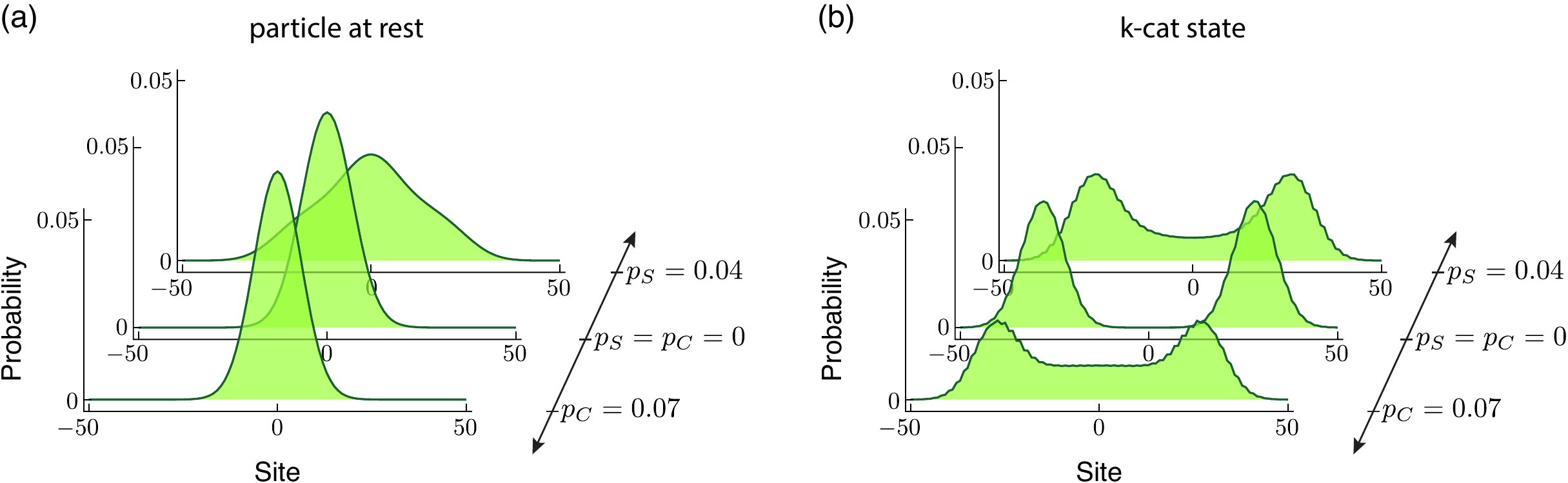}
	\caption{Effect of decoherence on spatial wave packets with narrow Gaussian quasi momentum distributions after a 40-step Hadamard quantum walk. Initially, a wave packet with a RMS width $\Delta x_0 =8/\sqrt{2}$ sites is been prepared in the upper band in (a) $k=0$ and in a superposition of $k=\pm \pi/2$.}
	\label{fig:Figure7}
\end{figure*}

Rather than considering infinitely delocalized eigenstates $\ket{s_\pm(k)}\otimes\ket{k}$ in our decoherence model, we opt for realistic, Gaussian-shaped wave packets $\ket{s_\pm(k)}\otimes \sum_{x}\exp(-x^2/(4\Delta x_0^2))\exp(i k x)$, which constitute an excellent approximation of the walk's eigenstates in the limit of large $\Delta x_0$.
The analysis of the spatial probability distribution presented in Figure~\ref{fig:Figure7} allows us to grasp the principal differences between spin and spatial decoherence.

Note first, though, that without decoherence the walker at rest preserves its position centered at $x=0$, though its RMS size increases according to $\Delta x(t)\approx \sqrt{\Delta x_0^2+n^2/(2m^\star\Delta x_0)^2}$ \cite{Jackson:1962}, with $m^\star=\pm|\tan(\theta/2)|$ being the effective mass in the upper ($+$) and lower ($-$) band at $k=0$; the size of the walker thus remains essentially unchanged for $n<m^\star\Delta x_0^2$ steps before expanding ballistically with $n$.
In contrast, the walker in the $k$-cat state splits into two counterpropagating wave packets, which propagate at the maximum group velocity, $\pm \cos(\theta/2)$. The bimodal coherent distribution reflects the two lateral peaks appearing in the quantum walk distribution of an initially localized particle, see Figure~\ref{fig:Figure4}.

Considering the effect of decoherence, we notice in Figure~\ref{fig:Figure7} that the spatial distribution of the $k{=}0$ wave packet is nearly unaffected by spin decoherence, while it is significantly dispersed by spatial decoherence.
For the $k$-cat state, in contrast, the spatial distribution is dispersed for both types of decoherence and it exhibits a noticeable occupation of the central positions \textemdash although, importantly, without exhibiting any central peak.
These two observations indicate that the cusp-like peak occurring in the presence of spin decoherence, see Figure~\ref{fig:Figure4}, indeed originates  from the quasi momentum components near $k=0$.

\begin{figure*}[b]
	\centering
		\includegraphics[width=.9\textwidth]{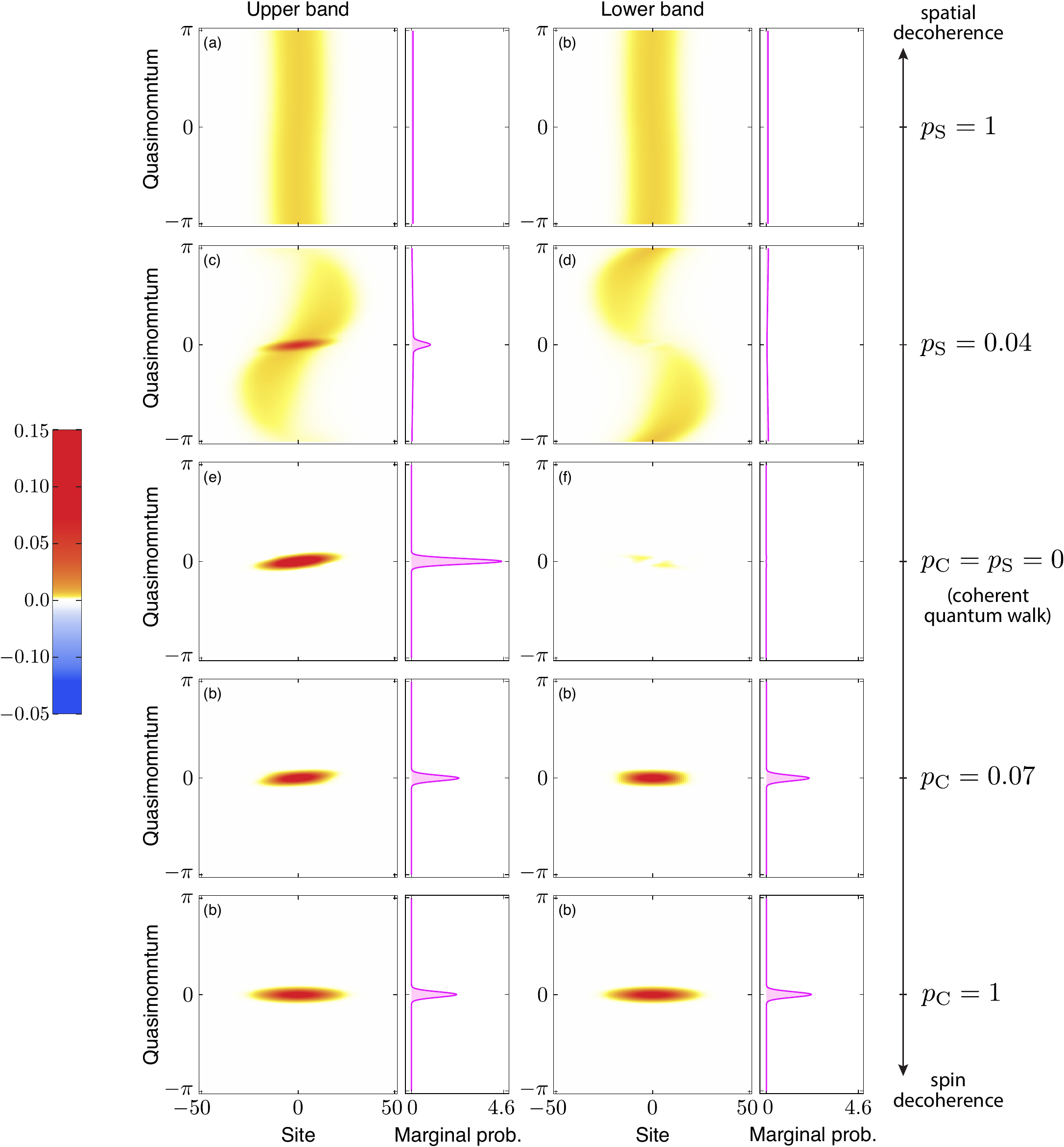}
\caption{Decoherence of a narrow $k$-distribution around $k{=}0$: Wigner functions and quasi momentum marginals are displayed after 40 steps for different amounts of spin and spatial decoherence. The slightly skewed orientation of the Wigner profile in the central row is due to the non-vanishing spread in $k$-space of the initial wave packet.}
	\label{fig:Figure8}
\end{figure*}

\begin{figure*}[b]
	\centering
		\includegraphics[width=.9\textwidth]{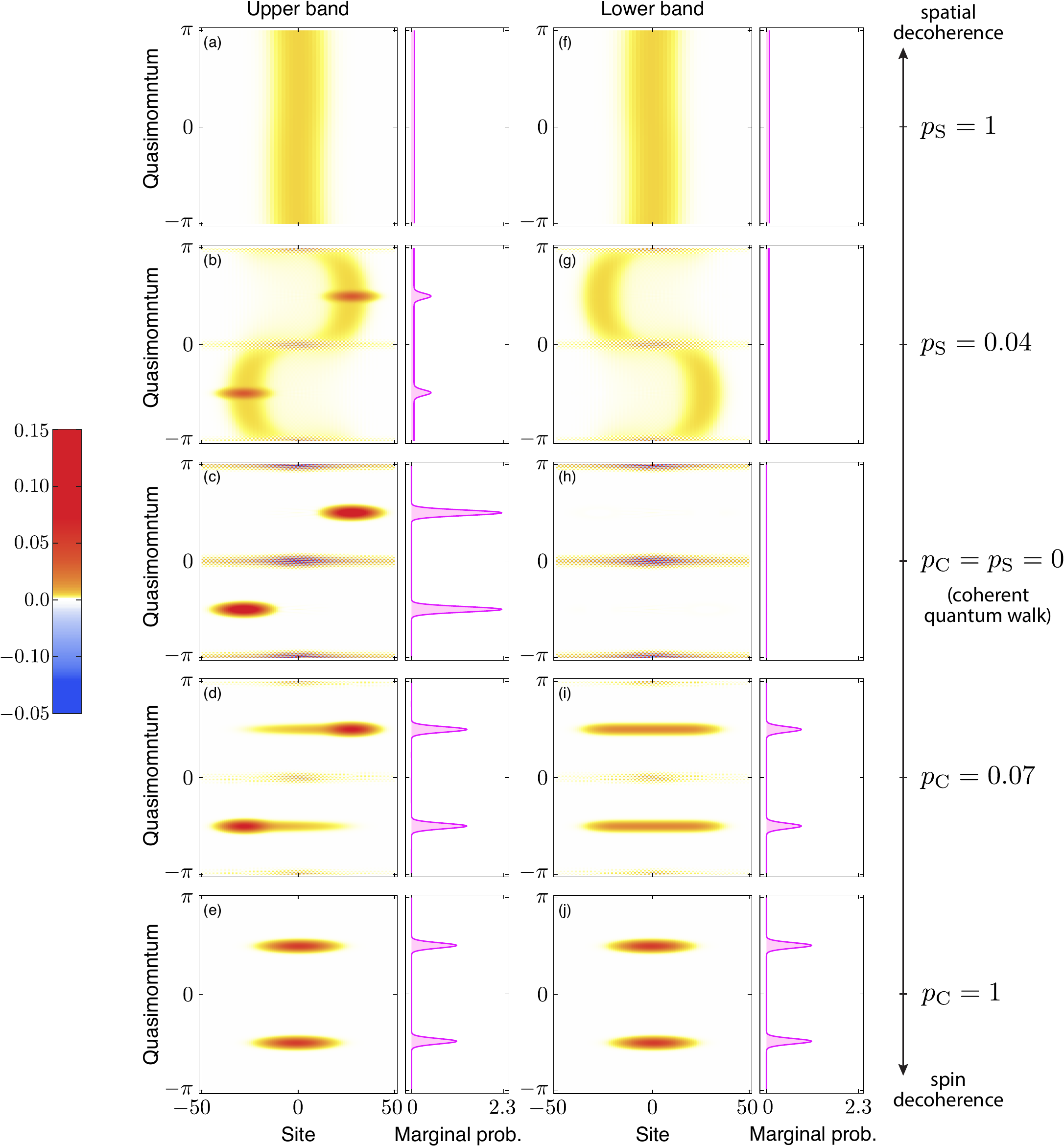}
	\caption{Decoherence of a coherent bimodal $k$-distribution around $k{=}\pm\pi/2$: Wigner functions and quasi momentum marginals are displayed after 40 steps for different amounts of spin and spatial decoherence.}
	\label{fig:Figure9}
\end{figure*}

The Wigner function enables a visual representation and a more detailed comparison of the influence of position and spin decoherence on the coherence properties in $k$-space.
The two limiting cases presented in Figure~\ref{fig:Figure7} are considered separately in Figure~\ref{fig:Figure8} and Figure~\ref{fig:Figure9}, which illustrate their phase space dynamics.
In essence, the main effects are:
\begin{itemize}
\item Spin decoherence causes a repopulation of the two bands without modifying the $k$-distributions.
\item Spatial decoherence causes collapse of the wavefunction into individual lattice sites. Hence, each decoherence event is associated with a rapid spreading in $k$-space.
\end{itemize}

\subsubsection{Decoherence of a walker at rest.}

The Wigner function in the absence of decoherence (central row of Figure~\ref{fig:Figure8}) shows that the walker retains the initial phase-space distribution nearly intact after 40 steps: namely, a minimal uncertainty Gaussian-shaped profile populating a single band only.
Owing to the large position uncertainty, $\Delta x_0\gg1$, the quasi momentum is narrowly spread within the Brillouin zone, $\Delta k_0 = 1/(2\Delta x_0)$. 

Spin decoherence leads to an incoherent and even population of the two bands while leaving the $k$-space distribution unaffected. The latter is a consequence of the translational invariance of the Kraus operators $\hat{E}_s$ in Equation~(\ref{eq:krausops}).
Hence, spin decoherence does not significantly modify the spatial distribution for a small number of steps $n<\Delta x_0^2$, because both bands exhibit around $k=0$ the same motional dynamics of a particle at rest.
Note also that the narrow momentum distribution in the Brillouin zone implies spatial coherences extending over several lattice sites, which survive in spite of spin decoherence \textemdash even for maximum decoherence rate $p_\text{C}=1$.

Spatial decoherence, instead, leads to partial projection of the wave packet onto individual lattice sites. Thus, the $k$-space volume is evenly filled, making large group velocity components available to the walker. Because of that, mild spatial decoherence can even accelerate the initial expansion of the wave packet, see also Figure~\figref{fig:Figure7}{a}.

\subsubsection{Decoherence of $k$-cat states.}

Like in the foregoing example, the Wigner function in the absence of decoherence (central row of Figure~\ref{fig:Figure9}) is concentrated in a single band.
The two outbound peaks move at the maximum possible speed of the quantum walk, preserving the original, narrow momentum spread.
In addition, two thin, horizontal stripes of the Wigner function exhibit negative values near $k=0$ and $k=\pi$, which are regarded as the evidence of the non-classicality of the $k$-cat state.
This behavior resembles the phase-space Schr\"odinger cats generated with a micromaser set-up \cite{Deleglise:2008}.

Spin decoherence populates the complementary band without changing the $k$-distribution.
For equal $k$-values, the complementary band group velocity is opposite to the original group velocity.
This stochastic changes of direction prevent the gap between the two outbound peaks to form.
Multiple band-changing processes tend to equilibrate the spatial distribution in the original band towards a Gaussian profile, too.
The vanishing of the negative-valued stripes indicates the rapid loss of coherence between the two outbound peaks.
Spatial decoherence leaves a fraction of incoherent population behind, which in turn starts a new secondary quantum walk and again prevents the gap around $x=0$.

\section{Decoherence Model of Dephasing Processes with Long Time Memory}\label{sec:long-time-memory}

The effects of decoherence strongly reflect whether the decoherence process has a memory of the previous history of the walk.
In case there is no memory, decoherence process is said to be Markovian, which is the scenario discussed in Section~\ref{sec:phenomomdel}, while in case some memory of the past is retained, decoherence is said to be non-Markovian, since in this case the environment or the perturbing external field exhibits non-zero correlations over a timescale comparable to (or even longer than) the duration of the walk.

Non-Markovian decoherence of discrete-time quantum walks has hitherto been studied for the particular case of a one-step-long memory \cite{McGettrick:2010}.
Ahlbrecht \emph{et al.}~\cite{Ahlbrecht:2011} developed methods based on perturbation theory to compute the long-time asymptotic spatial distributions.
Here, we provide an exact solution for quasi-stationary spin dephasing processes, which are of great interest for the experiments.
We specifically consider the situation in which the walk operator $\hat{W}$ is modified to
\begin{equation}
	\hat{W}_\zeta=\exp(i\zeta\hat{\sigma}_z/2)\hat{W}\,,
\end{equation} 
which accounts for a quasi energy shift $\zeta$ between the two spin states.
In our model, we assume $\zeta$ to be constant for the entire duration of the walk, and to be randomly drawn on each realization of the walk with a certain probability distribution $f(\zeta)$.
This model is, in fact, suited to describe shot-to-shot fluctuations of the energy difference between the two spin components, which can be caused by slowly varying external perturbations or inhomogeneous (i.e., different for each walker) energy shifts, see also Section~\ref{sec:decoh_mech}.

The key property of this decohered quantum walk consists in the fact that its time evolution is computed using the same walk operator $\hat{W}$ but displaced in $k$-space by $\zeta/2$, $\hat{W}_\zeta(k)=\bra{k}\hat{W}_\zeta(k)\ket{k}=\hat{W}(k-\zeta/2)$.
Thus, the quasi energy spectrum in Equation~(\ref{eq:bands}) is also displaced by the same amount, $\omega_{\zeta,\pm}(k)=\omega_\pm(k-\zeta/2)$.
The spatial probability distribution of the walker averaged over $\zeta$ can be exactly derived for a walker prepared initially in a single site $\ket{x_0}$ with spin $\ket{\sigma}$.
The probability amplitude of finding the walker after $n$ steps at position $x$ with spin $s$ is
\begin{equation}\label{eq:walkzeta}
	a_\zeta(x,s)=\bra{s,x}\hat{W}_\zeta^n\ket{\sigma,x_0}
	=e^{-i(x-x_0)\zeta/2} \bra{s,x}\hat{W}^n\ket{\sigma,x_0}\,,
\end{equation}
where the identity $\sum_{\pm,k}\ket{s_{\pm}(k-\zeta/2)}   \bra{s_{\pm}(k-\zeta/2)}\otimes\ket{k}\bra{k}$ has been inserted in front of $\ket{\sigma,x_0}$.
One notable aspect of this decoherence process is that the spatial probability distribution $\hspace{-2pt}\int\hspace{-1pt}\mathrm{d}\zeta\hspace{2pt} f(\zeta) \hspace{2pt} |a_\zeta(x,s)|^2 = |\bra{s,x}\hat{W}^n\ket{\sigma,x_0}|^2$ is equal to that of a fully coherent walk. Hence, neither the spatial distribution nor the local spin orientation do depend on $\zeta$, and the expansion remains ballistic.
We therefore conclude that the effect of such a dephasing mechanism is not observable through local quantum-state tomography techniques \cite{Karski:2009}.
Note that the spatial probability distribution of a wave packet with defined initial quasi momentum $k$ does, on the contrary, depend on $\zeta$.
In that case, the wave packet splits into two components of different weights because of the two bands, with their positions being displaced after $n$ steps by $v_{\text{g},\pm}(k-\zeta/2)\hspace{0.5pt}n$.
The final distribution can be obtained by averaging over $\zeta$.

\begin{figure*}[t]
	\centering
		\includegraphics[width=.85\textwidth]{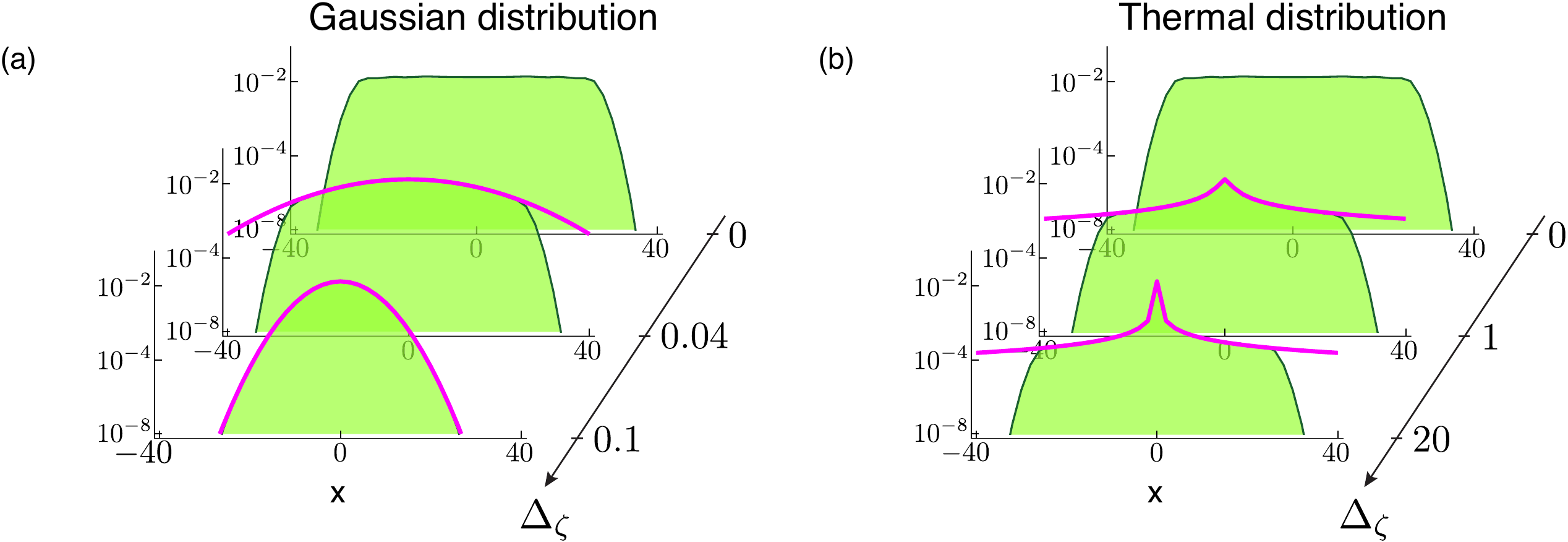}
	\caption{Single particle correlation function $|G(x,-x)|$ of a 40-step Hadamard quantum walk with localized initial state (like in Figure~\ref{fig:Figure5}). The shaded green curves are obtained by numerically computing the evolution of the quantum walker for increasing values of the inhomogeneous dephasing $\Delta_\zeta$ in ase of (a) a Gaussian distribution and of (b) a thermal Boltzmann distribution in a two-dimensional harmonic potential. The computed curves are identical to those predicted by the analytic formula in Equation~(\ref{eq:spatcohinhom}). The magenta thick lines display the integral term in the analytic formula, which represents the amount of coherence suppression.
}
	\label{fig:Figure10}
\end{figure*}

Unlike the probability distribution, the spatial coherences of an initially localized walker do depend on $\zeta$, and become suppressed by averaging over the distribution  $f(\zeta)$.
Using the probability amplitudes in Equation~(\ref{eq:walkzeta}), the single-particle correlation function introduced in (\ref{eq:singlepartcorrfunc}) is obtained by computing
\begin{equation}\label{eq:spatcohinhom}
G(x,y) = G_0(x,y) \hspace{-2pt}\int\hspace{-1pt}\mathrm{d}\zeta\hspace{2pt} f(\zeta) \hspace{2pt}  e^{-i(y-x)\zeta/2}\,.
\end{equation}
The result puts the correlation function of the decohered walk directly in relation with that one of the unperturbed walk.
The integral yields a suppression the spatial correlations (that is, the spatial coherences), which is worth computing for two special distributions: For a Gaussian distribution, $f(\zeta)=(2\pi \Delta_\zeta^2)^{-1/2}\exp[-\zeta^2/(2\Delta_\zeta^2)]$ with $-\infty<\zeta<\infty$, we obtain $|G(x,y)| = |G_0(x,y)| \exp[-\Delta_\zeta^2 \hspace{2pt}(x-y)^2/2]$, which exhibits a Gaussian suppression of spatial coherences as a function of the distance between two sites, as shown in Figure~\figref{fig:Figure10}{a}.

In the second example, we consider the distribution $f(\zeta)=\exp[-\zeta/\Delta_\zeta]/\Delta_\zeta$ with $\zeta>0$,
which represents in dimensionless units the potential energy distribution of a thermal atom in a two-dimensional harmonic trap \cite{Kuhr:2005}.
This distribution is physically relevant for our system because it describes the differential light shift experienced by an atom cooled to the ground state of the longitudinal motion in the optical lattice (cf.~the discussion about differential light shifts in Appendix~\ref{sec:appendix}).
We obtain the correlation function
\begin{equation}\label{eq:coh_length_thermal}
	|G(x,y)| = \frac{|G_0(x,y)|}{\sqrt{1+\Delta_\zeta^2\hspace{2pt}(x-y)^2/4}}\,,
\end{equation}
which is shown in Figure~\figref{fig:Figure10}{b}.
We define the coherence length $\ell=2/\Delta_\zeta$ as the characteristic spatial scale on which coherences extend, as shown in Equation~(\ref{eq:coh_length_thermal}).
The remarkable finding is that the coherence length $\ell$ can directly be expressed in terms of the inhomogeneous coherence time $T_2$ as defined in Ref.~\cite{Kuhr:2005}.
We obtain $\ell=T_2/\tau$, where $\tau$ is the duration of a single step.
This result permits an intuitive interpretation of the coherence length in terms of the number of steps that can be performed within the coherence time.

\section{Physical Decoherence Mechanisms in Neutral Atom Experiments}\label{sec:decoh_mech}

A classification scheme that provides insight into decoherence of discrete-time quantum walks relies on two criteria, which distinguish
\begin{enumerate}
	\item{}spatial and spin decoherence,
	\item{}whether the decoherence mechanism is directly induced by the environment independently of the walk operations (environment-induced decoherence) or, instead, it modifies the behavior of the coin operation (coin-mediated decoherence) or the shift operation (shift-mediated decoherence).
\end{enumerate}
Using these criteria, Table~\ref{tab:decoherence_mechanisms} reports the decoherence mechanisms that affect discrete-time quantum walks in neutral atom experiments. In Appendix~\ref{sec:appendix}, we discuss each decoherence mechanism individually. We also provide an estimate of the coherence length $\ell$ and of the decoherence rates $p_\text{C}$ and $p_\text{S}$ for those mechanisms that can be described using the decoherence model developed in this work.

\begin{table}
\centering
\newif\ifpdftable
\pdftabletrue
\ifpdftable
\includegraphics{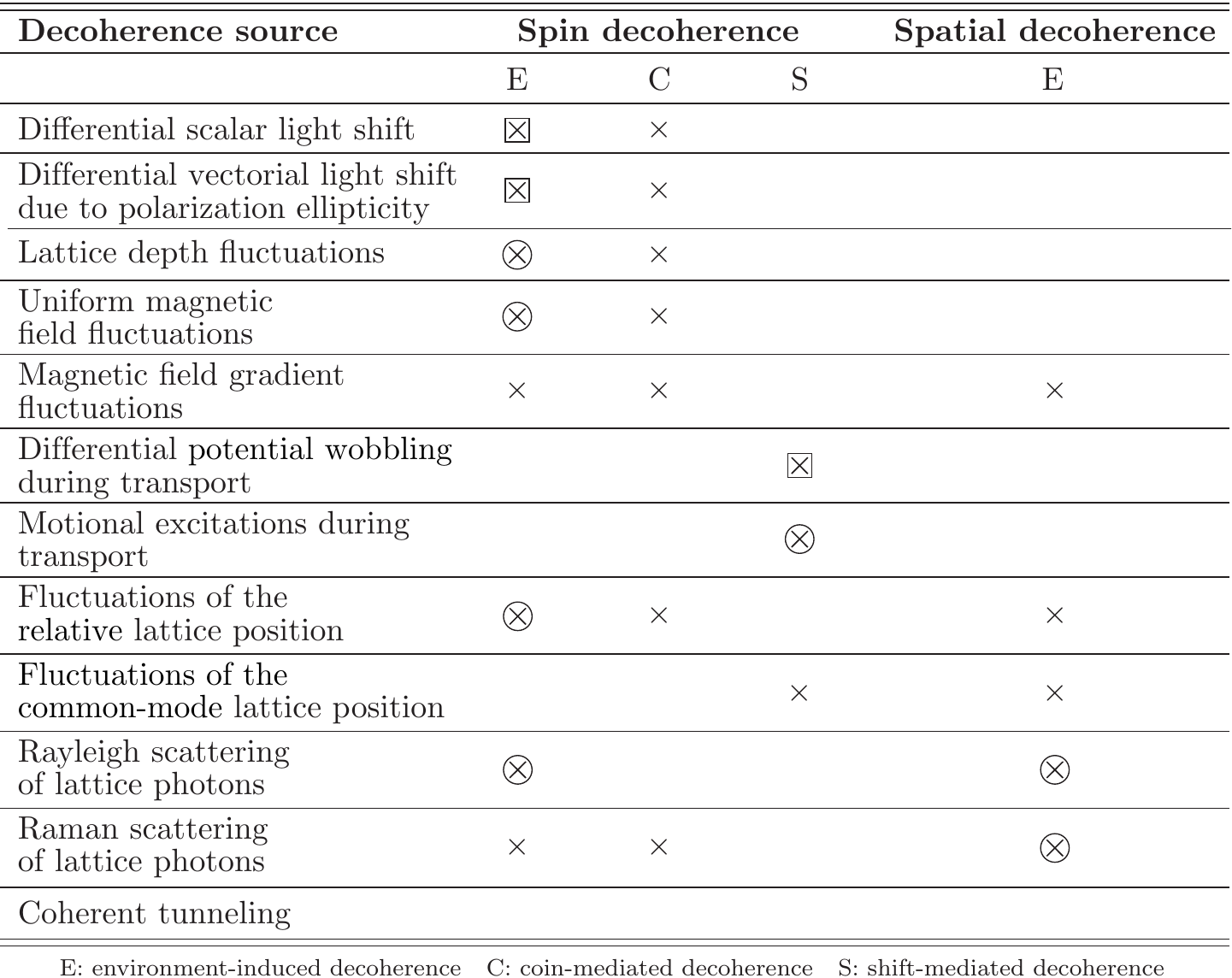}
\else
{\newsavebox{\withcircle}
\savebox{\withcircle}{$\times$\hspace{-7mm}
\setlength{\unitlength}{1mm}
\begin{picture}(4.3, 0)\put(2.6,1.08){\circle{3.9}}\end{picture}}
\small\newlength{\firstcol}\setlength{\firstcol}{4.6cm}
\begin{tabular}{p{\firstcol}cccc}
\hline
\hline
	\textbf{Decoherence source} & \multicolumn{3}{l}{\textbf{Spin decoherence}} & {\textbf{Spatial decoherence}} \\
\hline\\[-4mm]
& \hspace{5mm}E\hspace{5mm} & \hspace{5mm}C\hspace{5mm} & \hspace{5mm}S\hspace{5mm} & \hspace{5mm}E\hspace{5mm}  \\
\hline\\[-4mm]
\parbox{\firstcol}{Differential scalar\\[-1mm] light shift} & \usebox{\withcircle} & $\times$ & \\[2mm]
\hline\\[-4.mm]
{Lattice depth fluctuations} & \usebox{\withcircle} & $\times$ & \\[0.5mm]
\hline\\[-4.mm]
\parbox{\firstcol}{Uniform magnetic\\[-1mm]field fluctuations} & \usebox{\withcircle} & $\times$ &  \\[2mm]
\hline\\[-4.mm]
\parbox{\firstcol}{Magnetic field gradient\\[-1mm] fluctuations} & $\times$ & $\times$ & & $\times$ \\[2mm]
\hline\\[-4.mm]
\parbox{\firstcol}{Jitter of\\[-1mm] lattice laser polarization} & $\times$  & $\times$ & \\[2mm]
\hline\\[-4.mm]
\parbox{\firstcol}{Jitter of\\[-1mm] lattice position} & & & $\times$ & $\times$ \\[2mm]
\hline\\[-4.mm]
\parbox{\firstcol}{Rayleigh scattering\\[-1mm] of lattice photons} & & & & \usebox{\withcircle} \\[2.mm]
\hline\\[-4.mm]
\parbox{\firstcol}{Raman scattering\\[-1mm] of lattice photons} & $\times$ & $\times$ & & \usebox{\withcircle} \\[2.mm]
\hline
\hline
\multicolumn{5}{l}{\scriptsize{}E: environment-induced decoherence\quad{}C: coin-mediated decoherence\quad{}S: shift-mediated decoherence}
\end{tabular}}
\fi
\caption{\label{tab:decoherence_mechanisms}Dominant physical decoherence mechanisms affecting quantum walks of single atoms in state-dependent optical lattices. Translationally invariant mechanisms contribute to spin decoherence. Mechanisms that do not resolve the spin state but break the translational symmetry contribute to spatial decoherence. Circled crosses set apart decoherence mechanisms that are accounted for by either the $\hat{E}_{s}$ or $\hat{E}_{x}$ operator in Equation~(\ref{eq:krausops}), and boxed crosses identify quasi-stationary dephasing effects discussed in Section~\ref{sec:long-time-memory}. See Appendix~\ref{sec:appendix} for a detailed discussion of each mechanism.}
\end{table}

\section{Discussion and outlook}

In this work we have covered two classes of decoherence mechanisms, which either affect the internal or external degrees of freedom of discrete-time quantum walks.
We have shown that our phenomenological decoherence model is able give a precise account of the measured spatial distributions, which we recorded for quantum walks up to about 100 steps in our neutral atom experiment.
For Markov decoherence mechanisms, the spin and spatial decoherence rates, which are defined as the amount of decoherence per step, give us quantitative, accurate indicators of the coherence of the quantum walk.
In our experiment, for instance, we routinely employ these rates for the fine tuning of the experimental apparatus.

We have separately considered decoherence mechanisms with memory. An analytic solution for the case of quasi-stationary spin dephasing, which is especially relevant for real experiments, has been obtained.
Unexpectedly, we find that the local properties of the quantum walk are not affected by this decoherence mechanism.
Spatial coherences, however, decay over a finite coherence length, which we express as a function of the dephasing strength.
Future experimental investigations should be able to measure the coherence length, e.g., with atom interferometry techniques and compare it with the theoretical prediction.

We should also note that, in spite of the specific form of the Kraus operators, which have been chosen to describe our experiment, several of our findings apply, in general, to the whole class of either spin or spatial decoherence:
In fact, the repopulation of all momentum states by spatial decoherence and, \emph{vice versa}, the preservation of the momentum distribution by spin decoherence are a consequence of the translational symmetry of the respective Kraus operators, rather than depending on specific details.

The discrete Wigner function representation developed in this work provides a novel tool for the visualization of decoherence phenomena in discrete-time quantum walks.
This representation has proven useful to relate the spectral properties of discrete-time quantum walks in momentum space to the effects induced by decoherence.
We have shown how spin and spatial decoherence, which exhibit striking differences in the momentum distribution, leave a visible imprint in the position distribution as well, namely, the central cusp-like maximum.
The direct experimental investigation of quantum walks in the momentum space would provide further insight into decoherence mechanisms.
To that end, wave packets with defined quasi momentum, that is with a coherent delocalization over several lattice sites, need to be prepared by either sub-recoil cooling of atomic motion \cite{Ammann:1997} or by directly using a Bose-Einstein condensate.

Finally, we remark that a measurement of the Wigner function represents an interesting experimental avenue to explore, which would enable direct experimental access to the phase space properties of quantum walks.
Further theoretical investigation is, however, needed to identify measurement schemes that permit an efficient reconstruction of the Wigner function.
One possibility could be provided by the measurement of the expectation value of the displaced parity operator~\cite{Royer:1977,Leibfried:1996,Banaszek:1999,Deleglise:2008}:  spatial displacements can be realized with translations of the state-dependent optical lattice, while momentum displacements can be produced by means of an external artificial electric field~\cite{Genske:2013}. A method to measure the parity of the wavefunction remains to be found.

\section{Acknowledgments}
We thank Andreas Steffen and Maximilian Genske for fruitful discussions and for their experimental contribution during an early stage of this work.
We also thank Carsten Robens and Stefan Brakhane for insightful discussions.
We acknowledge the contribution of Jai-Min Choi for the experimental characterization of inhomogeneous dephasing.
We acknowledge the financial support of NRW-Nachwuchsforschergruppe ``Quantenkontrolle auf der Nanoskala'', ERC grant DQSIM, EU project SIQS.

\section{References}
\ifusebibfile

\bibliographystyle{njpbibstyle}
\bibliography{decohbib}

\providecommand{\newblock}{}
\begin{thebibliography}{10}
\expandafter\ifx\csname url\endcsname\relax
  \def\url#1{{\tt #1}}\fi
\expandafter\ifx\csname urlprefix\endcsname\relax\def\urlprefix{URL }\fi
\providecommand{\eprint}[2][]{\url{#2}}

\bibitem{Blatt:2012}
Blatt R and Roos C~F 2012 {\em Nature Phys.\/} {Quantum simulations with
  trapped ions} \href{http://dx.doi.org/10.1038/nphys2252}{{\bf 8} 277}

\bibitem{Devoret:2013}
Devoret M~H and Schoelkopf R~J 2013 {\em Science\/} {Superconducting circuits
  for quantum information: an outlook.}
  \href{http://dx.doi.org/10.1126/science.1231930}{{\bf 339} 1169}

\bibitem{Karski:2009}
Karski M, F{\"o}rster L, Choi J~M, Steffen A, Alt W, Meschede D and Widera A
  2009 {\em Science\/} {Quantum Walk in Position Space with Single Optically
  Trapped Atoms} \href{http://dx.doi.org/10.1126/science.1174436}{{\bf 325}
  174}

\bibitem{Kendon:2007}
Kendon V 2007 {\em Math. Struct. Computer Science\/} Decoherence in quantum
  walks -- a review \href{http://dx.doi.org/10.1017/S0960129507006354}{{\bf 17}
  1169}

\bibitem{Dur:2002}
D{\"u}r W, Raussendorf R, Kendon V~M and Briegel H~J 2002 {\em Phys. Rev. A\/}
  {Quantum walks in optical lattices}
  \href{http://dx.doi.org/10.1103/PhysRevA.66.052319}{{\bf 66} 052319}

\bibitem{Brun:2003a}
Brun T~A, Carteret H~A and Ambainis A 2003 {\em Phys. Rev. Lett.\/} {Quantum to
  Classical Transition for Random Walks}
  \href{http://dx.doi.org/10.1103/PhysRevLett.91.130602}{{\bf 91} 130602}

\bibitem{Annabestani:2010}
Annabestani M, Akhtarshenas S~J and Abolhassani M~R 2010 {\em Phys. Rev. A\/}
  {Decoherence in a one-dimensional quantum walk}
  \href{http://dx.doi.org/10.1103/PhysRevA.81.032321}{{\bf 81} 032321}

\bibitem{Ahlbrecht:2011}
Ahlbrecht A, Vogts H, Werner A~H and Werner R~F 2011 {\em Journal of
  Mathematical Physics\/} {Asymptotic evolution of quantum walks with random
  coin} \href{http://dx.doi.org/10.1063/1.3575568}{{\bf 52} 042201}

\bibitem{Ahlbrecht:2012}
Ahlbrecht A, Cedzich C, Matjeschk R, Scholz V~B, Werner A~H and Werner R~F 2012
  {\em Quantum Information Processing\/} {Asymptotic behavior of quantum walks
  with spatio-temporal coin fluctuations}
  \href{http://dx.doi.org/10.1007/s11128-012-0389-4}{{\bf 11} 1219}

\bibitem{Kendon:2003}
Kendon V and Tregenna B 2003 {\em Phys. Rev. A\/} {Decoherence can be useful in
  quantum walks} \href{http://dx.doi.org/10.1103/PhysRevA.67.042315}{{\bf 67}
  042315}

\bibitem{Broome:2010}
Broome M~A, Fedrizzi A, Lanyon B~P, Kassal I, Aspuru-Guzik A and White A~G 2010
  {\em Phys. Rev. Lett.\/} {Discrete Single-Photon Quantum Walks with Tunable
  Decoherence} \href{http://dx.doi.org/10.1103/PhysRevLett.104.153602}{{\bf
  104} 153602}

\bibitem{Schreiber:2011}
Schreiber A, Cassemiro K~N, Poto{\v c}ek V, G{\'a}bris A, Jex I and Silberhorn
  C 2011 {\em Phys. Rev. Lett.\/} {Decoherence and Disorder in Quantum Walks:
  From Ballistic Spread to Localization}
  \href{http://dx.doi.org/10.1103/PhysRevLett.106.180403}{{\bf 106} 180403}

\bibitem{Genske:2013}
Genske M, Alt W, Steffen A, Werner A~H, Werner R~F, Meschede D and Alberti A
  2013 {\em Phys. Rev. Lett.\/} {Electric Quantum Walks with Individual Atoms}
  \href{http://dx.doi.org/10.1103/PhysRevLett.110.190601}{{\bf 110} 190601}

\bibitem{Schmitz:2009}
Schmitz H, Matjeschk R, Schneider C, Glueckert J, Enderlein M, Huber T and
  Schaetz T 2009 {\em Phys. Rev. Lett.\/} {Quantum Walk of a Trapped Ion in
  Phase Space} \href{http://dx.doi.org/10.1103/PhysRevLett.103.090504}{{\bf
  103} 090504}

\bibitem{Zahringer:2010}
Zaehringer F, Kirchmair G, Gerritsma R, Solano E, Blatt R and Roos C~F 2010
  {\em Phys. Rev. Lett.\/} {Realization of a Quantum Walk with One and Two
  Trapped Ions} \href{http://dx.doi.org/10.1103/PhysRevLett.104.100503}{{\bf
  104} 100503}

\bibitem{robens:2014}
Robens C, Alt W, Meschede D, Emary C and Alberti A 2015  {Ideal Negative
  Measurements in Quantum Walks Disprove Theories Based on Classical
  Trajectories} (\textit{Preprint}
  \eprint[http://arxiv.org/abs/1404.3912]{arXiv:1404.3912 [quant-ph]})

\bibitem{Ivanov:2008}
Ivanov V~V, Alberti A, Schioppo M, Ferrari G, Artoni M, Chiofalo M~L and Tino
  G~M 2008 {\em Phys. Rev. Lett.\/} {Coherent Delocalization of Atomic Wave
  Packets in Driven Lattice Potentials}
  \href{http://dx.doi.org/10.1103/PhysRevLett.100.043602}{{\bf 100} 043602}

\bibitem{Belmechri:2013}
Belmechri N, F{\"o}rster L, Alt W, Widera A, Meschede D and Alberti A 2013 {\em
  J. Phys. B: At. Mol. Phys.\/} {Microwave control of atomic motional states in
  a spin-dependent optical lattice}
  \href{http://dx.doi.org/10.1088/0953-4075/46/10/104006}{{\bf 46} 104006}

\bibitem{note:TOF}
{The momentum distribution could be measured with a time-of-flight measurement,
  which is best suited for large atomic ensembles rather than single atoms.}

\bibitem{Steffen:2012}
Steffen A, Alberti A, Alt W, Belmechri N, Hild S, Karski M, Widera A and
  Meschede D 2012 {\em Proc. Natl. Acad. Sci. USA\/} {Digital atom
  interferometer with single particle control on a discretized space-time
  geometry.} \href{http://dx.doi.org/10.1073/pnas.1204285109}{{\bf 109} 9770}

\bibitem{note:continuousWalks}
{For continuous-time systems, mild and strong decoherence regimes are
  distinguished depending on whether the decoherence rate is smaller or bigger
  than the tunneling rate, respectively.}

\bibitem{Haroche:2006}
Haroche S and Raimond J~M 2006 {\em {Exploring the Quantum: Atoms, Cavities,
  and Photons}\/} (New York: {Oxford University Press})

\bibitem{Berry:1977}
Berry M~V 1977 {\em Philos. Trans. R. Soc. London, Ser. A\/} {Semi-Classical
  Mechanics in Phase Space: A Study of Wigner's Function}
  \href{http://dx.doi.org/10.1098/rsta.1977.0145}{{\bf 287} 237}

\bibitem{Bizarro:1994}
Bizarro J~P 1994 {\em Phys. Rev. A\/} {Weyl-Wigner formalism for rotation-angle
  and angular-momentum variables in quantum mechanics.}
  \href{http://dx.doi.org/10.1103/PhysRevA.49.3255}{{\bf 49} 3255}

\bibitem{Hinarejos:2012}
Hinarejos M, P{\'e}rez A and Banuls M~C 2012 {\em New J. Phys.\/} {Wigner
  function for a particle in an infinite lattice}
  \href{http://dx.doi.org/10.1088/1367-2630/14/10/103009}{{\bf 14} 103009}

\bibitem{Hinarejos:2013}
Hinarejos M, Banuls M~C and P{\'e}rez A 2013 {\em J. Comput. Theor. Nanosci.\/}
  {A Study of Wigner Functions for Discrete-Time Quantum Walks}
  \href{http://dx.doi.org/10.1166/jctn.2013.3101}{{\bf 10} 1626}

\bibitem{Bizarro:2013}
Bizarro J~P~S 2013 {\em New J. Phys.\/} {Comment on {\textquoteleft}Wigner
  function for a particle in an infinite lattice{\textquoteright}}
  \href{http://dx.doi.org/10.1088/1367-2630/15/6/068001}{{\bf 15} 068001}

\bibitem{Monroe:1996}
Monroe C, Meekhof D~M, King B~E and Wineland D~J 1996 {\em Science\/}
  {A``Schrodinger Cat'' Superposition State of an Atom}
  \href{http://dx.doi.org/10.1126/science.272.5265.1131}{{\bf 272} 1131}

\bibitem{Jackson:1962}
Jackson J~D 1962 {\em Classical Electrodynamics\/} (New York: John Wiley \&
  Sons) {Chapter 7}

\bibitem{Deleglise:2008}
Del{\'e}glise S, Dotsenko I, Sayrin C, Bernu J, Brune M, Raimond J~M and
  Haroche S 2008 {\em Nature\/} {Reconstruction of non-classical cavity field
  states with snapshots of their decoherence}
  \href{http://dx.doi.org/10.1038/nature07288}{{\bf 455} 510}

\bibitem{McGettrick:2010}
McGettrick M 2010 {\em Quantum Inf. Comput.\/} One dimensional quantum walks
  with memory {\bf 10} 509

\bibitem{Kuhr:2005}
Kuhr S, Alt W, Schrader D, Dotsenko I, Miroshnychenko Y, Rauschenbeutel A and
  Meschede D 2005 {\em Phys. Rev. A\/} {Analysis of dephasing mechanisms in a
  standing-wave dipole trap}
  \href{http://dx.doi.org/10.1103/PhysRevA.72.023406}{{\bf 72} 023406}

\bibitem{Ammann:1997}
Ammann H and Christensen N 1997 {\em Phys. Rev. Lett.\/} {Delta Kick Cooling: A
  New Method for Cooling Atoms}
  \href{http://dx.doi.org/10.1103/PhysRevLett.78.2088}{{\bf 78} 2088}

\bibitem{Royer:1977}
Royer A 1977 {\em Phys. Rev. A\/} {Wigner function as the expectation value of
  a parity operator} \href{http://dx.doi.org/10.1103/PhysRevA.15.449}{{\bf 15}
  449}

\bibitem{Leibfried:1996}
Leibfried D, Meekhof D~M, King B~E, Monroe C, Itano W~M and Wineland D~J 1996
  {\em Phys. Rev. Lett.\/} {Experimental Determination of the Motional Quantum
  State of a Trapped Atom}
  \href{http://dx.doi.org/10.1103/PhysRevLett.77.4281}{{\bf 77} 4281}

\bibitem{Banaszek:1999}
Banaszek K, Radzewicz C, W{\'o}dkiewicz K and Krasi{\'{n}}ski J~S 1999 {\em
  Phys. Rev. A\/} {Direct measurement of the Wigner function by photon
  counting} \href{http://dx.doi.org/10.1103/PhysRevA.60.674}{{\bf 60} 674}

\bibitem{Steffen:2013}
Steffen A, Alt W, Genske M, Meschede D, Robens C and Alberti A 2013 {\em Rev.
  Sci. Instrum.\/} {Note: In situ measurement of vacuum window birefringence by
  atomic spectroscopy.} \href{http://dx.doi.org/10.1063/1.4847075}{{\bf 84}
  126103}

\bibitem{Kim:2013}
Kim H, Han H~S and Cho D 2013 {\em Phys. Rev. Lett.\/} {Magic Polarization for
  Optical Trapping of Atoms without Stark-Induced Dephasing}
  \href{http://dx.doi.org/10.1103/PhysRevLett.111.243004}{{\bf 111} 243004}

\bibitem{deChiara:2008}
de~Chiara G, Calarco T, Anderlini M, Montangero S, Lee P~J, Brown B~L, Phillips
  W~D and Porto J~V 2008 {\em Phys. Rev. A\/} {Optimal control of atom
  transport for quantum gates in optical lattices}
  \href{http://dx.doi.org/10.1103/PhysRevA.77.052333}{{\bf 77} 052333}

\bibitem{Negretti:2013}
Negretti A, Benseny A, Mompart J and Calarco T 2013 {\em Quantum Information
  Processing\/} {Speeding up the spatial adiabatic passage of matter waves in
  optical microtraps by optimal control}
  \href{http://dx.doi.org/10.1007/s11128-012-0357-z}{{\bf 12} 1439}

\bibitem{Cedzich:2013}
Cedzich C, Ryb{\'a}r T, Werner A~H, Alberti A, Genske M and Werner R~F 2013
  {\em Phys. Rev. Lett.\/} {Propagation of Quantum Walks in Electric Fields}
  \href{http://dx.doi.org/10.1103/PhysRevLett.111.160601}{{\bf 111} 160601}

\bibitem{Vacchini:2009}
Vacchini B and Hornberger K 2009 {\em Physics Reports\/} {Quantum linear
  Boltzmann equation}
  \href{http://dx.doi.org/10.1016/j.physrep.2009.06.001}{{\bf 478} 71}

\bibitem{grimm:2000}
Grimm R, Weidem{\"u}ller M and Ovchinnikov Y 2000 {\em Adv. At. Mol. Opt.
  Phy.\/} Optical dipole traps for neutral atoms
  \href{http://dx.doi.org/10.1016/S1049-250X(08)60186-X}{{\bf 42} 95}

\bibitem{Cline:1994}
Cline R~A, Miller J~D, Matthews M~R and Heinzen D~J 1994 {\em Opt. Lett.\/}
  {Spin relaxation of optically trapped atoms by light scattering}
  \href{http://dx.doi.org/10.1364/OL.19.000207}{{\bf 19} 207}

\bibitem{Uys:2010}
Uys H, Biercuk M~J, Vandevender A~P, Ospelkaus C, Meiser D, Ozeri R and
  Bollinger J~J 2010 {\em Phys. Rev. Lett.\/} {Decoherence due to Elastic
  Rayleigh Scattering}
  \href{http://dx.doi.org/10.1103/PhysRevLett.105.200401}{{\bf 105} 200401}

\end{thebibliography}

\else

\providecommand{\newblock}{}

\fi

\begin{appendices}
\section{Detailed analysis of physical decoherence mechanisms}
\label{sec:appendix}

This appendix provides a thorough discussion of the physical decoherence mechanisms presented in Table~\ref{tab:decoherence_mechanisms}.
					 
\paragraph{Differential light shifts} displace the resonance frequency separating the two spin states by an amount proportional to the light intensity.

Atoms that are thermally distributed \textemdash like in our experiment, in which atoms have been cooled to the motional ground state only in the lattice direction \textemdash experience different values of the light intensity, which depends on their exact location inside the potential wells of the optical lattice.
Because the timescale of the transverse motion is on the order of $30$ steps, the	 motion along the transversal direction can be considered, for simplicity's sake, as frozen during the entire walk.
During the quantum walk, thus, each atom experiences a different detuning of the resonance frequency (inhomogeneous broadening), which produces a dephasing of the two spin components and also impairs the coin operation.

Each run of the quantum walk is characterized by a quasi energy shift between the two spin components (see Section~\ref{sec:long-time-memory}) that amounts to
$\zeta= |\eta|\hspace{2pt} U  \hspace{1pt}\tau /\hbar$, where $\eta$ denotes the relative differential light shift, and $U$ represents the potential depth at the position of the atom in the lattice.
We recall that $\tau$ is the duration of a single step of the walk.
The parameter $\eta$ takes into account scalar and vectorial components, $\eta=\eta_\text{s}+\eta_\text{v}'\epsilon$ with $\epsilon$ being the degree of ellipticity of the light polarization.
The ellipticity  $\epsilon=(I_+-I_-)/(I_++I_-)$ reflects the unbalance between the $I_\pm$ intensity components with $\pm$-circular polarization.

The scalar effect was modeled by Kuhr \emph{et al.}~\cite{Kuhr:2005} and is caused by the nonzero hyperfine frequency splitting $\Delta_\text{HF}$ between $\ket{\uparrow}$ and $\ket{\downarrow}$ states.
This effect yields 
\begin{equation}
	\eta_\text{s}=\Delta_\text{HF} \left(\frac{3}{2\Delta_\text{D1}+\Delta_\text{D2}}-\frac{1}{\Delta_\text{D1}}-\frac{1}{\Delta_\text{D2}}\right)\,.
\end{equation}
Here, $\Delta_\text{D1}$ and $\Delta_\text{D2}$ are the lattice detunings from the D1 and D2 lines.

The vectorial effect was modeled by Steffen \emph{et al.}~\cite{Steffen:2013} and is caused by imperfections of the light polarization (ellipticity) of the optical lattice.
This effect yields
\begin{equation}
	\eta_\text{v}'=\big[{m_F(\uparrow)} {g_F(\uparrow)}- {m_F(\downarrow)} {g_F(\downarrow)}\big] \hspace{3pt}\frac{\Delta_\text{D1}-\Delta_\text{D2}}{2\Delta_\text{D1}+\Delta_\text{D2}}\,.
\end{equation}
Here, $m_F(s)$ and $g_F(s)$ are, respectively, the magnetic quantum number and the $g$-factor associated with the $s$ spin state. Note that in the electronic ground state of alkali atoms $g_F(\downarrow)=-g_F(\uparrow)$.

In our experiment, with $g_F(\uparrow)=1/4$ we obtain $\eta_\text{s} =\num{2.5e-3}$ and $\eta'_\text{v}=7/4$.
Using the definition in Ref.~\cite{Kuhr:2005} of the inhomogeneous coherence time, $T_2=2\hbar/(|\eta| k_\text{B} T)$, we estimate that $T_{2}\approx\SI{600}{\micro\second}$ due to the scalar effect and $T_{2}\approx\SI{1}{\micro\second}/|\epsilon|$ for the vectorial effect, where the experimental transverse temperature $T_\text{2D}\approx \SI{10}{\micro\kelvin}$ has been used.
Based on the decoherence model in Section~\ref{sec:long-time-memory}, we compute the coherence length $\ell$ as the ratio $T_2/\tau$, obtaining
$\ell\approx \num{20}$ sites for the scalar effect and $\ell\approx\num{0.03}/\epsilon$ sites for the vectorial effect.
We measure the ellipticity caused by stress-induced birefringence in the vacuum cell employing the atoms themselves as a measurement probe \cite{Steffen:2013}.
We find typical values on the order of $\epsilon\approx \num{e-2}$ and correspondingly $\ell\approx\num{3}$ sites.
We showed in Ref.~\cite{Steffen:2013} that the optical birefringence can be  significantly suppressed, and thus the coherence length extended, by aligning the light polarization to one of the optical birefringence axes.
In addition, Kim \emph{et al.}~\cite{Kim:2013} showed that the vectorial component of the differential light shift can be exploited to compensate for the scalar one.

\paragraph{Fluctuations of the lattice depth} originate from intensity fluctuations and beam pointing instabilities of the optical lattice laser beams.
We distinguish common-mode intensity fluctuations of the two circularly polarized components,  $I_\text{cm}=(I_++I_-)/2$, from relative variations, $I_\text{r}=I_+-I_-$.
Common-mode intensity fluctuations make the scalar differential light shift vary in time,
while relative intensity fluctuations produce a vectorial differential light shift.
Providing time correlations across subsequent steps can be neglected (i.e., for sufficiently flat fluctuation spectra), the decoherence rate  $p_\text{C}=1-\mathcal{C}$ can be expressed in terms of the contrast $\mathcal{C}$ recorded by a Ramsey interferometer (with no spin echo) subject to the same fluctuation spectrum \cite{Kuhr:2005}. Recall that the contrast of a Ramsey interferometer is defined as twice the off-diagonal density matrix element of the qubit.
We can therefore compute $p_\text{C}$ in terms of the spectral power density of the external fluctuating fields.

Common-mode intensity fluctuations yield a spin decoherence rate $p_\text{C}=1-\exp(-\Delta\Phi^2/2)\approx \Delta\Phi^2/2$, where the phase variance is determined by the relative intensity noise spectrum $\mathrm{RIN}(\omega)$ of $I_\text{cm}$, 
\begin{equation}
	\Delta \Phi^2=\frac{\tau^2\hspace{1pt}\eta^2 \hspace{1pt}U_0^2}{\hbar^2} \hspace{-1pt}\int_0^\infty \hspace{-4pt}\mathrm{d}\omega\hspace{3pt} \mathrm{sinc}^2(\omega \tau/2)\hspace{2pt}\mathrm{RIN}(\omega),
\end{equation}
with $U_0$ being the potential depth at the bottom of the lattice.
In the integral, a low-frequency cut-off at around $1/t_\text{tot}$ is implicitly assumed, which accounts for the finite overall duration, $t_\text{tot}$, of a quantum walk experiment.

Relative intensity fluctuations, instead, translate into a time-varying polarization ellipticity, $\epsilon=I_\text{r}/(2I_\text{cm})$, which produces a vectorial differential light shift.
In this case, the phase variance determining $p_\text{C}$ is related to the noise spectral density $S_{\epsilon}(\omega)$ of the ellipticity,
\begin{equation}
	\Delta \Phi^2=\frac{\tau^2\hspace{0.3pt}\eta_\text{v}'\hspace{-1pt}{}^2
\hspace{1pt}U_0^2}{\hbar^{2}} \hspace{-1pt}\int_0^\infty \hspace{-4pt}\mathrm{d}\omega\hspace{3pt} 
\mathrm{sinc}^2(\omega \tau/2)\hspace{2pt}S_{\epsilon}(\omega).
\end{equation}
Because $\eta_\text{v}'\gg\eta$, relative fluctuations are potentially more harmful than common-mode fluctuations of the same magnitude.
Fortunately, the magnitude of relative fluctuations is expected to be small since these fluctuations can only be produced by the electro-optic modulator through mechanical resonances, thermal gradients or voltage noise.

\paragraph{Uniform magnetic field fluctuations} produce a homogeneous displacement of the resonance frequency.
Like for fluctuations of the lattice depth, we can estimate the spin decoherence rate as $p_\text{C}\approx\Delta\Phi^2/2$,
where the phase variance is determined by the noise spectral density $S_B(\omega)$ of the magnetic field $B$ along the quantization axis (namely, the lattice direction), \begin{equation}
\Delta\Phi^2=\frac{\tau^2\hspace{0.3pt}\mu_B^2}{\hbar^2}\hspace{1pt} [{m_F(\uparrow)} {g_F(\uparrow)}-{m_F(\downarrow)} {g_F{(\downarrow)}}]^2
\hspace{-3pt}\int_0^\infty\hspace{-8pt} \mathrm{d}\omega\hspace{3pt} 
\mathrm{sinc}^2(\omega \tau/2)\hspace{2pt}S_{B}(\omega).
\end{equation}
Experimentally, we directly measure the loss of contrast due to magnetic field fluctuations by performing a microwave Ramsey spectroscopy with free-falling atoms, which are therefore not perturbed by the optical dipole trap.
The measurement shows that the contrast drops to about $\SI{50}{\percent}$ at around $20\,\tau$.
Modeling the loss of contrast with an exponential curve, we estimate a decoherence rate $p_\text{C}$ between $\SI{3}{\percent}$ and $\SI{4}{\percent}$,
which would correspond to $\Delta \Phi \approx \num{7e-2}$, or equivalently to a magnetic field white noise $S_B(\omega)\approx (\SI{4.5e-6}{\gauss/\sqrthertz})^2/(2\pi)$ in the bandwidth determined by $1/\tau\approx \SI{30}{\kilo\hertz}$.
In reality, the noise spectrum is not flat, but rather peaked around the power line frequency ($\SI{50}{\hertz}$) and its harmonics.
To minimize power line disturbances, the quantum walk is performed in the experiment synchronously with the power line signal.

\paragraph{Magnetic field gradient fluctuations} are conceptually similar to uniform magnetic field fluctuations.
Their contribution is expected to be smaller than that of uniform magnetic fields since the size of a quantum walk (a few tens of microns) is several orders of magnitude smaller than the distance separating the atoms from the magnetic field sources.
In addition to spin decoherence, a fluctuating field gradient also contributes to spatial decoherence since it brings about position-dependent dephasing.

\paragraph{Differential potential wobbling} refers to the spin-dependent variation of the optical lattice depth, which occurs during the shift operation \cite{Belmechri:2013}.
During the transport of the atom, in fact, the optical lattice changes from a lin-$\parallel$-lin to lin-$\angle$-lin configuration, until it reaches again a lin-$\parallel$-lin configuration.
The largest vectorial differential light shift is experienced when the lattice is in a lin-$\perp$-lin configuration.
In this configuration, the relative differential light shift between the two spin components amounts to
\begin{equation}
	\eta_\perp=\frac{{|m_F(\uparrow)}\hspace{-0.5pt}|\hspace{3pt} {g_F(\uparrow)}+ {|m_F(\downarrow)\hspace{-0.5pt}|}\hspace{3pt} {g_F(\downarrow)}}{2} \hspace{1pt}\frac{\Delta_\text{D1}-\Delta_\text{D2}}{2\Delta_\text{D1}+\Delta_\text{D2}}\,.
\end{equation}
Following similar arguments as in the foregoing discussion about differential light shift, we estimate the inhomogeneous coherence length $\ell$ using the model developed in Section~\ref{sec:long-time-memory}.
With $\eta_\perp=1/8$ in our experiment and approximating the average relative differential light shift during the shift operation with $\eta_\perp/2$, we find $\ell\approx 0.8$ sites.
In spite of the very short coherence length, shot-to-shot dephasing has no effect on the local properties of the quantum walk, as proven in Section~\ref{sec:long-time-memory}.
Evidence of this strong inhomogeneous dephasing due to differential potential wobbling is independently obtained in our experiment by atom interferometry along the lines of Ref.~\cite{Steffen:2012}, but omitting the spin echo.
We observe that the contrast drops dramatically down to about $\SI{30}{\percent}$ already when the atom's wavefunction is split to 1 lattice site separation.
The observed contrast drop is in accordance with our estimate of the coherence length.
In atom interferometry applications \cite{Steffen:2012}, this dephasing mechanism is suppressed by using refocussing spin echo pulses.

We remark that the coherence length can be extended by performing a feedforward modulation of the $I_+$ and $I_-$ intensity components, which should be chosen such as to cancel out the potential wobbling. 
Alternatively, by cooling the atom to the three-dimensional ground state, we expect to fully suppress this dephasing mechanism, since it vanishes for atoms in the three-dimensional ground state.
As a third alternative, differential potential wobbling can be fully avoided by properly choosing the internal states defining the coin Hilbert space.
For instance, the pair of states  $\ket{\downarrow}=\ket{F=3,m_F=3}$ and $\ket{\uparrow}=\ket{F=4,m_F=3}$ would result in $\eta_\perp=0$.

\paragraph{Motional excitations during transport} make the atom oscillate inside the potential well.
Motional excitations are represented by coherent Glauber states, which resemble the motion of a classical harmonic oscillator.
Because of the differential potential wobbling during transport (see the foregoing discussion regarding differential potential wobbling), motional excitations can differ for the two spin components.
After a shift operation, the two spin components may thus oscillate with a different phase and amplitude.
Consequently, the overlap between the wave packets of each spin component is reduced, and the interference is partially suppressed.
The suppression of interference can effectively be described in terms of spin dephasing.
This decoherence effect can be suppressed by minimizing (for instance, by means of optimal control theory \cite{deChiara:2008,Negretti:2013}) the motional excitations for both spin components.
A precise evaluation of the decoherence rate $p_\text{C}$ requires further investigations, and it strongly depends on the specific transport details.

For our choice of coin states, the depth of the lattice potential experienced by the $\ket{\uparrow}$ component remains unchanged during the whole shift operation, while the depth of the lattice potential experienced by the $\ket{\downarrow}$ component undergoes a depth modulation, which reaches $3/4$ in relative units at the midpoint of the transport sequence \cite{Belmechri:2013}.
We therefore expects more excitations to occur for the $\ket{\downarrow}$ state than for the $\ket{\uparrow}$ state.
For both spin states, we measured the motional excitations produced after one single step with microwave sideband spectroscopy \cite{Belmechri:2013}.
From that measurement, we estimate an upper limit for the motional excitations between $\SI{1}{\percent}$ and $\SI{2}{\percent}$ per shift operation.
Based on these values, we can confidently estimate that  $p_\text{C}<\SI{2}{\percent}$.

\paragraph{Fluctuations of the lattice position} can be of two types: (1) fluctuations of the relative distance between the two optical lattices acting, respectively, on the $\ket{\uparrow}$ and $\ket{\downarrow}$ spin components, (2) common-mode fluctuations of the average position of the two optical lattices.
Relative position fluctuations are caused by jittering of the polarization angle of one of the two linearly polarized laser beams forming the optical lattice \cite{Belmechri:2013}.
Polarization jitter can occur, for instance, because of mechanical resonances inside the electro-optic modulator.
Common-mode fluctuations, instead, are caused by mechanical vibrations of the optical elements (e.g., mirrors) employed in the optical lattice set-up.

In both cases, position fluctuations impair spatial coherences by exerting fluctuating inertial forces.
A fluctuating inertial force imprints a phase gradient along the lattice.
The phase gradient experienced by the two spin components is opposite in sign for relative fluctuations, whereas it is identical for common-mode fluctuations, providing the atom is at rest.
During the shift operation, however, the effect of common-mode fluctuations can differ for the two spin components.

Conceptually, the effect of position fluctuations on the quantum walk is similar to that of external fluctuating electric field acting on a charged particle in a crystal.
The theory developed for static electric quantum walks \cite{Cedzich:2013} provides the mathematical framework to account for slow fluctuations of the common-mode position.

Furthermore, relative position fluctuations cause variations of the Franck-Condon factors for both the microwave carrier and the motional sideband transitions \cite{Belmechri:2013}.
The consequences of this effect are twofold: on the one hand, fluctuations of the carrier's Franck-Condon factor make the microwave Rabi frequency fluctuate in time, thus impairing the coin operation.
On the other hand, fluctuations of the sidebands' Franck-Condon factors at the longitudinal vibrational frequency can induce resonant excitations of higher motional states during the coin operation.
Both effects are experimentally mitigated by letting mechanical resonances in the electro-optic modulator damp down for a sufficiently long time (a few microseconds) after executing the shift operation and before applying the coin operation.

Lastly, relative position fluctuations can cause motional excitations  differing for the two spin components.
Spin-dependent motional excitations contribute to spin dephasing (see previous discussion of motional excitations during transport).
We expect this effect to be negligible since the measured timescale necessary for a single excitation from the ground state to the first excited state is about $\SI{1}{\second}$, which is several orders of magnitude longer than the single step time $\tau$.

\paragraph{Spontaneous scattering of lattice photons} has a twofold effect: (1) it destroys spatial coherences by projecting the quantum walker's wave packet into a single lattice site, and (2) it also causes decoherence of the spin degree of freedom.

Concerning spatial coherences, the spatial decoherence rate per step, $p_\text{S}$, is defined as the probability that a scattering event occurs in a time $\tau$. For a detailed treatment of collisional decoherence of matter waves, see, for instance, \cite{Vacchini:2009}.
By calling $\Gamma_\text{tot}$ the overall scattering rate, it results that $p_\text{S}=\Gamma_\text{tot}\tau$.
If we neglect the hyperfine splitting of the atomic ground and excited states, the expression of $\Gamma_\text{tot}$ for an alkali atom is found in the literature \cite{grimm:2000},
\begin{equation}\label{eq:totalscattering}
	\Gamma_\text{tot}=\alpha
	\left(\frac{2}{\Delta_\text{D2}^2}+\frac{1}{\Delta_\text{D1}^2}\right),
\end{equation}
where 
$\alpha=(\omega_L/\omega_\text{D1})^{3}\hspace{1pt}{\Gamma_\text{D1}\hspace{1pt} \Omega_R^2}/{12}\approx (\omega_L/\omega_\text{D2})^{3}\hspace{1pt}{\Gamma_\text{D2}\hspace{1pt} \Omega_R^2}/{12}$ with $\Gamma_{\text{D1}}$,$\Gamma_{\text{D2}}$ and $\omega_\text{D1},\omega_\text{D2}$ being, respectively, the natural linewidths and the transition frequencies of the D1 and D2 lines (the two definitions of $\alpha$ are equivalent if one neglects small relativistic corrections).
The Rabi frequency $\Omega_R=\mathcal{E}_0\mu/\hbar$ is defined in terms of the largest dipole moment element $\mu=|\bra{ J'{=}3/2,m_J'{=}3/2} \vec{d}\hspace{1pt} \ket{J{=}1/2,m_J{=}1/2}|$ and the electric field amplitude.
For our experimental parameters, we obtain $\Gamma_\text{tot}=\SI{14}{\second^{-1}}$ and $p_\text{S}\approx\num{4e-4}$.

Concerning the effect of scattering on the spin variable, we distinguish two types of scattering depending on whether the atom remains in the same internal state (elastic process, Rayleigh scattering) or not (inelastic process, Raman scattering). The total scattering rate comprises both elastic and inelastic processes, $\Gamma_\text{tot}=\Gamma_\text{el}(s)+\Gamma_\text{inel}(s)$ with $s$ identifying the spin state $\ket{s}$.

Raman scattering induces decoherence by mixing the spin populations.
For an atom with nuclear spin $I\neq 0$ and assuming the rotating wave approximation, we compute the inelastic scattering rate $\Gamma_\text{inel}$ using Fermi's golden rule and the Wigner-Eckart theorem.
The result depends on the polarization of the optical lattice's electric field, which is in our case linearly polarized and, importantly, orthogonal to the quantization axis, which is set along the lattice direction (for linear polarization along the quantization axis see, e.g., Ref.~\cite{Cline:1994}).
We obtain
\begin{equation}
	\Gamma_\text{inel}(s) = \alpha
		\hspace{1pt}
		\frac{2-g_F(s)^2m_F(s)^2}{3}
\left(\frac{1}{\Delta_\text{D2}}-\frac{1}{\Delta_\text{D1}}\right)^{\hspace{-1pt}2}.
\end{equation}
For large detunings, the formula shows that the Raman scattering rate (i.e., the rate of events flipping the electronic spin state) is much smaller than the total scattering rate of Equation~(\ref{eq:totalscattering}).
However, that is not the case when $\omega_L$ lies within D1 and D2 line like in the case of state-dependent optical lattices.
Among all Raman scattering events, those events that couple the two coin states, thus that remain within the coin Hilbert space, occur with the same rate for both $s$ states, which amounts to  $\Gamma_\text{inel}^\text{qubit}=I/(I+1/2)\Gamma_\text{inel}(\uparrow)$ (for simplicity's sake, the analytic formula is here provided only for the most relevant case when $g_F(\uparrow)m_F(\uparrow){=}-g_F(\downarrow)[m_F(\downarrow)+1]{=}1$).
For cesium atoms, the ratio $I/(I+1/2)=7/8$ shows that the majority of Raman events remain within the coin space. For our experiment, we obtain $\Gamma_\text{inel}(\uparrow)=\SI{5}{\second^{-1}}$ and $\Gamma_\text{inel}(\downarrow)=\SI{7}{\second^{-1}}$.

Raman scattering events that pump the atom's internal state outside the coin Hilbert state preclude the application of the coin operation in the subsequent steps since the atom's state is no longer resonant with the microwave radiation employed for the coin operation.
Raman scattering events that keep the atom's internal state inside the coin Hilbert space cause a relaxation of spin populations and spin coherences with a rate $\Gamma_\text{inel}^\text{qubit}$.
In both situations, an account of the effect of Raman scattering on the quantum walk requires developing a different decoherence model beyond pure dephasing.
The effect of Raman processes, however, is much smaller than that produced by other decoherence mechanisms because of the small scattering rates $\Gamma_\text{inel}(s)$.

In contrast to Raman processes, Rayleigh scattering events preserve spin populations and cause only pure spin dephasing.
The two spin states $\ket{s}$ scatter the lattice circularly-polarized photons (note that the lattice linear polarization orthogonal to the quantization axis is an equal superposition of both circular polarizations) with probability amplitudes that differ in phase and magnitude.
The elastic scattering, thus, reveals to the environment information about the internal spin state $s$, which is  stochastically measured at a rate $\Gamma_\text{el.deph.}<\Gamma_\text{el}$. By generalizing the result of Uys \emph{et al.}~\cite{Uys:2010} derived for a pure spin-$1/2$ system, we obtain for an alkali atom with $I\neq0$
\begin{equation}
	\Gamma_\text{el.deph.}=\alpha  \hspace{1pt}
	\frac{\big[g_F(\uparrow)\hspace{1.3pt}m_F(\uparrow)-g_F(\downarrow)\hspace{1.3pt}m_F(\downarrow)\big]^2}{6}
\left(\frac{1}{\Delta_\text{D2}}-\frac{1}{\Delta_\text{D1}}\right)^{\hspace{-1pt}2}.
\end{equation}
Elastic scattering therefore causes spin decoherence of the quantum walk with a spin decoherence per step $p_\text{C}=	\Gamma_\text{el.deph.}\tau$. For our experiment, we obtain $\Gamma_\text{el.deph.}=\SI{7}{\second^{-1}}$ and $p_\text{C}=\num{2e-4}$.

\paragraph{Coherent tunneling} allows the atom to cross the potential barrier of the optical lattice and to coherently interfere with the adjacent wavefunction.
In a strict sense, this effect cannot be considered as a decoherence mechanism, since it alters the property of the walk, though in a fully coherent way.
Tunneling is expected to play no role in our experiment, since its timescale is many orders of magnitude longer than the duration of the quantum walk.
A numerical study of incoherent tunneling in discrete-time quantum walks was considered in previous work \cite{Dur:2002}.

\end{appendices}

\end{document}